# Material Development and Physical Properties of $BiS_2$-Based Layered Compounds

Yoshikazu Mizuguchi*

*Department of Physics, Tokyo Metropolitan University, 1-1, Minami-osawa, Hachioji 192-0397, Japan*

**Abstract**

In 2012, we discovered superconductivity in $Bi_4O_4S_3$ and $LaO_{1-x}F_xBiS_2$, whose crystal structures are composed of alternate stacks of a $BiS_2$-type conduction layer and a blocking layer. Since the discovery, many related superconductors and functional materials have been synthesized. Furthermore, the possibility of unconventional superconductivity has been proposed in recent theoretical and experimental studies. In addition, notable correlations between local structure and physical properties have been revealed in the $BiS_2$-based systems. In this review article, we summarize the material development (probably all the compounds) and physical properties of $BiS_2$-related materials developed in the last six years. The key parameters for the emergence of bulk superconductivity in the $BiS_2$-based compounds are carrier doping to the parent phase (band insulator), pressure-induced structural transition, in-plane chemical pressure, and the suppression of in-plane disorder. We systematically review those parameters and the method of designing new superconductors with $BiS_2$ layers. In addition, studies of $BiS_2$-based compounds as thermoelectric materials are briefly reviewed.

## 1. Introduction

Since the discovery of Cu-oxide high-transition-temperature (high-$T_c$) superconductors [1-4], various kinds of layered compounds have been synthesized with expectations of the discovery of high-$T_c$ and/or unconventional superconductivity. Another layered high-$T_c$ superconductor family is the Fe-based superconductors [5-10]. These high-$T_c$ superconductors have a layered structure composed of alternate stacks of a superconducting (SC) layer ($CuO_2$ or FeAs layers) and an insulating (blocking: BL) layer. In such a layered superconductor family, new superconductors can be designed and synthesized by replacing the BL layer structure [11]. Owing to the richness of the structural variation and the exotic properties of the superconducting states, layered superconductors have been extensively studied in the fields of physics, chemistry, and engineering.

In 2012, layered superconductors with a $BiS_2$-type SC layer were discovered [12,13]. As the first member, the $BiS_2$-based compound $Bi_4O_4S_3$ was reported [12]. The crystal structure of the $Bi_4O_4S_3$ superconductor is composed of alternate stacks of the $BiS_2$ SC layer ($Bi_2S_4$ bilayer) and the $Bi_4O_4(SO_4)_{0.5}$ BL layer [Fig. 1(a)]. The stacking of a SC layer and a BL layer is similar to that of the Cu-oxide and Fe-based superconductors. By replacing the BL layer by a simpler REO layer (RE: rare earth), $REO_{1-x}F_xBiS_2$ superconductors [Fig. 1(d)] were discovered [13-20]. Furthermore, $BiS_2$-based superconductors with a BL layer different from the LaO layer have been synthesized, which will be introduced in detail in the next section. Recently, the various functionalities of thermoelectric, photocatalytic, and spintronics materials have been observed (or predicted) for layered chalcogenides related to the $BiS_2$-based superconductors. Therefore, for the further development of the layered metal chalcogenide family as high-$T_c$ and unconventional superconductors or promising functional materials, understanding the crystal structure and physical properties is crucial.

In this review article, we summarize the material development of the $BiS_2$-related layered compounds and their physical and chemical properties. First, material information from Refs. 12–219, categorized into properties shown in the paper, is summarized in Table I. On the basis of the table, the essence of the structural variation and the physical properties is described. Furthermore, we try to systematically clarify the physical and chemical parameters essential for the emergence of superconductivity in $BiS_2$-based compounds. Particularly, the relationships among the crystal structure, electronic states, and physical properties are revealed. In addition, we show recent advances in the material design and clarification of superconductivity mechanisms in the $BiS_2$-based systems.

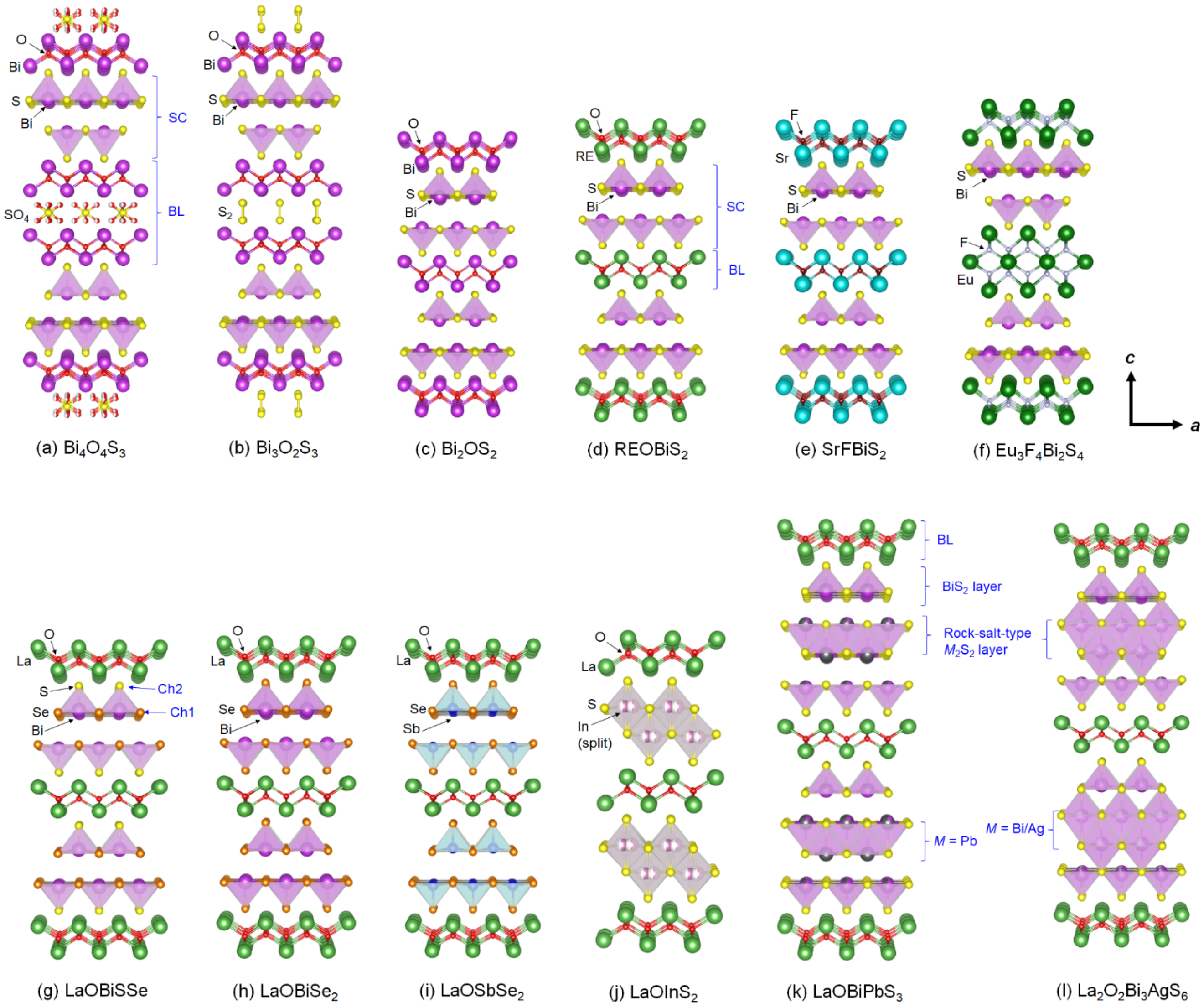

Fig. 1. (Color online) Schematic images of the crystal structure of $BiS_2$-based and related materials. The crystal structure images were prepared using VESTA software [220]. SC and BL denote superconducting and blocking layers, respectively.

**Table I. Material variation with references. Cp, AP, HP, HPA, HT, and HEA denote specific heat, ambient pressure, high pressure, HP annealing, high temperature, and high-entropy alloy, respectively.**

| **Materials** | **Maximum $T_c$ (K)** | Resistivity | Susceptibility | $C_p$ | Pressure effect | Single crystal | Band structure, theoretical study | Experiments for mechanisms, Spectroscopy, Local structure |
|---|---|---|---|---|---|---|---|---|
| $Bi_4O_4S_3$ | **6** | 12, 21, 22, 23, 24, 25, 26, 30, 36, 41, 54 | 12, 21, 22, 23, 25, 26, 31, 38, 41 | 25 | 24, 26, 41, 54 | | 12, 48 | 21, 30, 31, 41 |
| $(Bi,Ag)_4O_4S_3$ | **6** | 27 | 27 | | | | | |
| $(Bi,Mn)_4O_4S_3$ | **6** | 42 | 42 | | | | | |
| $(Bi,Ni)_4O_4S_3$ | **6** | 43 | 43 | | | | | |
| $Bi_4O_4(S,Se)_3$ | **4.5** | 33 | 33 | | | | | |
| $Bi_3O_2S_3$ | **6** | 34, 160 | 28, 34, 160 | 34 | | | 28, 95, 160 | |
| $(Bi,Y)_3O_2S_3$ | **5** | 44 | 44 | | | | | |
| $Bi_6O_8S_5$ | **-** | 29, 32 | 29, 32 | | | | | |
| $Bi_2OS_2$ ($BiOBiS_2$) | **-** | 36, 37 | 28 | | | | 95, 100, 154 | 154 |
| $Bi(O,F)BiS_2$ | **5** | 35, 37, 40 | 35, 37, 40 | | 39 | | | |
| $LaOBiS_2$ | **3.5 (doped devise)** | 13, 80, 81, 84, 123(HT), 132, 151, 152, 162(HT) | | | 80, 81 | 132, 151, 152, 170 | 18, 19, 89, 90, 91, 92, 94, 103, 105, 108, 109, 110, 154, 170 | 154, 161, 170, 207 |
| $La(O,F)BiS_2$ | **3( AP) 11 (HP) 11.5 (HPA)** | 13, 20, 45, 50, 53, 55, 56, 59, 80, 81, 82, 83, 84, 123(HT), 132, 138, 176, 201, 203 | 13, 45, 20, 50, 53, 81, 82, 138 | 20, 201 | 13, 24, 50, 53, 55, 56, 63, 80, 81, 82, 83, 176 | 59, 63, 74, 132, 134, 172, 201, 203 | 18, 19, 49, 89, 90, 91, 92, 93, 97, 101, 102(HP), 104, 105, 109, 134, 173 | 72, 73, 83, 134, 140, 141, 164, 172, 177, 203, 206, 210 |
| $(La,Ti)OBiS_2$ | **2.5 (AP) 3.5 (HP)** | 52, 208 | 52 | | 208 | | | |
| $(La,Zr)OBiS_2$ | **2** | 52 | 52 | | | | | |
| $(La,Hf)OBiS_2$ | **2.5** | 52 | 52 | 52 | | | | |
| $(La,Th)OBiS_2$ | **3** | 52, 208 | 52 | 52 | 208 | | 99 | |
| $(La,Sm)(O,F)BiS_2$ | **5 (AP) 10.5 (HP)** | 125, 126, 149, 150, 153 | 125, 126, 150 | 125 | 149, 150 | | | |
| $(La,Y)(O,F)BiS_2$ | **3** | 122 | 122 | 122 | | | | |
| $LaOBi(S,Se)_2$ | **-** | 156(HT), 162(HT), 166(HT), 184 | | | | | 106, 110 | 166, 207 |
| $LaOBiSe_2$ | **-** | 175 | | | | | | |
| $La(O,F)Bi(S,Se)_2$ | **4** | 79, 128, 133, 184, 195, 201 | 79, 128, 133, 184, 195 | 201 | | 133, 168, 201 | 184 | 133, 141, 168, 184, 186, 195, 201 |
| $La(O,F)BiSe_2$ | **3.5 (AP) 6.5 (HP)** | 62, 87, 88, 116, 117, 121, 139, 175, 202 | 62 ,87, 116, 117, 139, 175 | 117 | 116, 121, 139 | 87, 88, 116, 117, 121, 200, 202 | 98 | 62, 200 |
| $(La,Ce)OBiSSe$ | **3** | 196 | 196 | | | | | |
| $La(O,F)(Bi,Pb)S_2$ | **5** | 204 | 204 | | | 204 | | |
| $La(O,F)(Bi,Cd)S_2$ | **3** | 188 | 188 | | | 188 | | |
| $(La,Mg)(O,F)BiS_2$ | **3** | 76 | | | | | | |
| $CeOBiS_2$ | **1.5 (AP)**<br>**4 (HP)** | 14, 132, 171, 189 | 132, 189 | 132 | 189 | 132, 171, 189, 209 | | 124, 161, 189, 209 |
| $Ce(O,F)BiS_2$ | **3 (AP)**<br>**8 (HP)** | 14, 15, 20, 55, 56, 59, 61, 78, 86, 189 | 14, 15, 20, 59, 61, 78, 136, 189 | 20 | 15, 55, 56, 78, 86 | 59, 136, 143 | 97 | 124, 130, 135, 143, 167, 189, 206 |
| $(Ce,Nd)(O,F)BiS_2$ | **5** | 127 | 64, 127 | | | | | 141, 155 |
| $Ce(O,F)Bi(S,Se)_2$ | **2.5** | 159 | 159 | | | | | |
| $(Ce,Pr)OBiS_2$ | **2.5** | 216 | 216 | | | 216 | | |
| $PrOBiS_2$ | **-** | | | | | | | |
| $Pr(O,F)BiS_2$ | **4 (AP)**<br>**7 (HP)** | 16, 20, 56, 66, 68, 111, 147 | 16, 20, 66, 68, 147 | 20 | 56, 68, 111 | 147, 217 | 97 | 206, 217 |
| $Nd(O,F)BiS_2$ | **5 (AP)**<br>**6.5 (HP)** | 17, 20, 46, 54, 56, 57, 58, 59, 67, 71, 85, 112, 113, 163, 176 | 17, 20, 46, 57, 58, 67, 85, 112, 113 | 46 | 54, 56, 176 | 57, 58, 59, 70, 71, 75, 112, 113, 118, 137, 163, 187, 192 | 70, 71, 97, 112 | 58, 70, 71, 75, 112, 113, 118, 137, 163, 187, 192, 206 |
| $Nd(O,F)Bi(S,Se)_2$ | **4.5** | 77 | 77 | | | | | |
| $Nd(O,F)(Bi,Sb)S_2$ | **5** | 77 | 77 | | | | | |
| $Nd(O,F)(Bi,Pb)S_2$ | **6** | 158 | 158 | | | | | 158 |
| $(Nd,Sm)(O,F)BiS_2$ | **6** | 127 | 127 | | | | | 141, 155 |
| $Sm(O,F)BiS_2$ | **-** | | 126 | | | 126 | | |
| $Yb(O,F)BiS_2$ | **5** | 20 | 20 | 20 | | | | |
| $SrFBiS_2$ | **-** | 47, 51, 119 | | | 119 | | 47, 97, 100 | 96 |
| $(Sr,RE)FBiS_2$ | **10 (HP)** | 51, 60, 65, 69, 114, 115 | 51, 60, 65, 115 | 51, 65, 115 | 69, 114 | | 165 | 165, 190, 198 |
| $(Sr,RE)FBi(S,Se)_2$ | **4** | | | | | | | 199 |
| $(Sr,RE)FBiSe_2$ | **4** | 169 | 169 | | 169 | | | |
| $(Ca,La)FBiSe_2$ | **4** | 182 | 182 | | | | | |
| $EuFBiS_2$ | **0.3 (AP)**<br>**9 (HP)** | 120, 144, 146, 148(HT) | 120, 144 | 120 | 144, 146 | | 120 | 120, 181 |
| $(Eu,RE)FBiS_2$ | **2 (AP)**<br>**10 (HP)** | 142, 145 | 145 | 145 | 142 | | | |
| $(Eu,RE)FBi(S,Se)_2$ | **4** | 174, 191 | 174, 191 | | | | | 183, 185 |
| $Eu_3F_4Bi_2S_4$ | **1.5 (AP)**<br>**10 (HP)** | 129, 131, 157 | 129, 131 | 129 | 131 | 197 | | 197 |
| $Eu_3F_4Bi_2(S,Se)_4$ | **3.5** | 157 | 157 | 157 | | | | |
| $(Eu,Sr)_3F_4Bi_2S_4$ | **10 (HP)** | 178, 179 | 179 | | 178 | | | 179, 211 |
| $(Eu,Sr)_3F_4Bi_2(S,Se)_4$ | **3** | 180 | 180 | | | | | 180 |
| $REOSbS_2$ | **-** | | 215 | | | 215 | 110 | |
| $RE(O,F)SbSe_2$ | **-** | 213 | | | | | 107, 110 | |
| $REOInS_2$ | **-** | | | | | | 218 | 218 |
| $LaOBiPbS_3$ | **-** | 193, 205 | | | | | 194 | 194 |
| $La_2O_2Bi_3AgS_6$ | **0.5** | 205, 214 | | | | | | |
| $RE(O,F)BiS_2$ | **5 (HEA)** | 212, 219 | 212, 219 | | | | | 219 |

## 2. Material Development

### *2.1 Bi-O-S phases*

The field of $BiS_2$-based superconductors and related compounds was opened with the discovery of superconductivity in $Bi_4O_4S_3$ [12]. The discovery was serendipitous. When studying Bi-O-Cu-S, we removed Cu from the quaternary composition and observed superconducting signals in the ternary Bi-O-S compounds. By testing many compositions and annealing conditions (maybe over 150 conditions), we finally concluded that the bulk superconducting phase is $Bi_4O_4S_3$.

From X-ray diffraction and structural analyses, the crystal structure was determined as that in Fig. 1(a). The crystal structure of the $Bi_4O_4S_3$ superconductor is composed of alternate stacks of a $BiS_2$ SC layer and the $Bi_4O_4(SO_4)_{0.5}$ BL layer. The space group of $Bi_4O_4S_3$ is tetragonal *I*4/*mmm* ($D_{4h}^{17}$, No. 139). Band calculation suggested that the conduction band is mainly composed of Bi 6p$_x$ and 6p$_y$ orbitals hybridized with S 3p orbitals; hence, the $BiS_2$ layer acts as a conducting layer. The electron carriers are generated by the $SO_4$ defects: the valence states can be described as $(Bi^{3+}{}_2O^{2-}{}_2)_2(SO_4{}^{2-})_{0.5}(Bi^{3+}S^{2-}{}_2)_2 \cdot e^-$, where $e^-$ denotes a doped electron. The temperature dependences of resistivity and magnetic susceptibility are shown in Fig. 2. $T_c$ for $Bi_4O_4S_3$ is 4–6 K (onset temperature exceeds 8 K) in the bulk, but the feature of a superconducting gap was observed up to 14 K by spectroscopy [21], which suggested the possibility of a higher $T_c$ in this system.

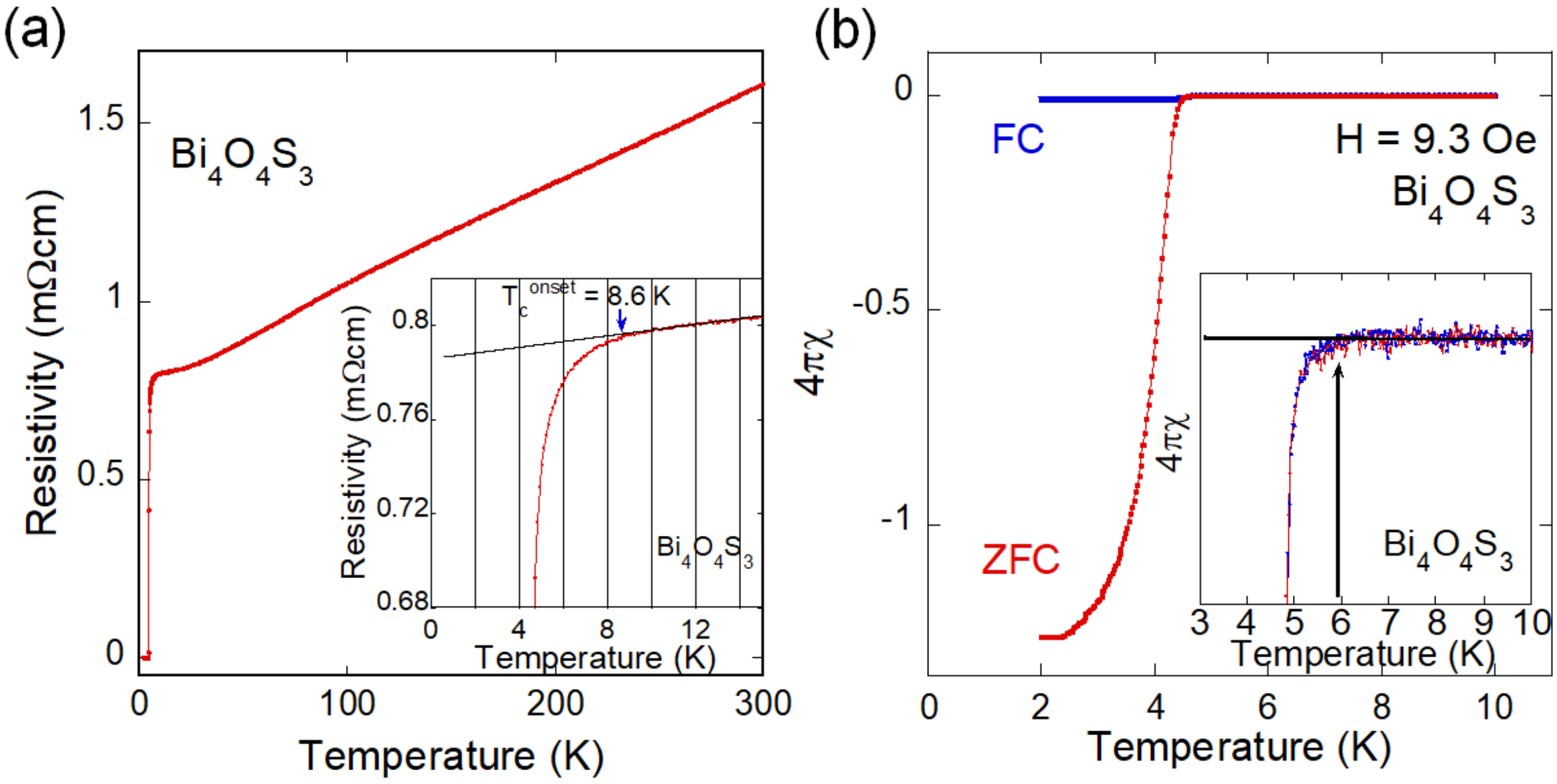

Fig. 2. (Color online) Temperature dependences of (a) electrical resistivity and (b) magnetic susceptibility of $Bi_4O_4S_3$. The original data was partly published in Ref. 12.

$Bi_3O_2S_3$ is a sister compound of $Bi_4O_4S_3$. As shown in Fig. 1(b), the $BiS_2$ SC layer is alternately stacked with the $Bi_4O_4S_2$ layer [28,34,160]. The space group of $Bi_3O_3S_2$ is tetragonal *I*4/*mmm*. The valence states of $Bi_3O_2S_3$ can be described as $(Bi^{3+}{}_2O^{2-}{}_2)_2S_2{}^{2-}(Bi^{3+}S^{2-}{}_2)_2$. The $Bi_3O_2S_3$ phase was suggested as a semimetal.

In the ternary Bi-O-S family, $Bi_2OS_2$ [36,37] has the simplest structure [Fig. 1(c)] similar to $LaOBiS_2$. The space group of $Bi_2OS_2$ is tetragonal *P*4/*nmm* ($D_{4h}^{7}$, No. 129). Although the parent phase is a band insulator [95,154], a partial substitution of O by F generates electron carriers, and the doped phases $BiO_{1-x}F_xBiS_2$ show superconductivity [35,37,40].

***2.2 REOBiS₂ type***

Among the $BiS_2$-based superconductors, the $REOBiS_2$-based phases have been mostly studied owing to their simple structure [Fig. 1(d)] and easy carrier doping control. The first report on the superconductivity in the $REOBiS_2$-based compound was for $LaO_{1-x}F_xBiS_2$ [13]. The parent phase ($x$ = 0) is a band insulator. F substitution makes the phase metallic, and superconductivity is observed in the doped phases. Later, we will explain in detail the carrier doping and crystal structure optimization for the emergence of bulk superconductivity in the $REOBiS_2$-based phases. Here, we summarize the variation of the $REOBiS_2$-based superconductors.

The crystal structure of $REOBiS_2$ is composed of $BiS_2$ conducting layers and REO BL layers. Almost all the $REOBiS_2$ and doped compounds have a tetragonal $P4/nmm$ structure. Exceptionally, $LaOBiS_2$ [161] and $LaOBiS_{2-x}Se_x$ [184 and this paper] undergo a structural transition from tetragonal to monoclinic $P2_1/m$ ($C_{2h}^2$, No. 11) at around room temperature. As listed in Table I, electron doping in $LaOBiS_2$ can be achieved by $F^-$ substitution for the $O^{2-}$ site or cation substitution (cation with a valence higher than 3+, such as $Ti^{4+}$) for the $La^{3+}$ site. In addition, by replacing La by Ce, Pr, Nd, Sm, or Yb, various $REOBiS_2$-based superconductors can be synthesized [14-17,20,126]. Also, the S site can be substituted by Se as displayed in Fig. 1(d) [62,79,159]. The doped Se tends to occupy the in-plane site in $LaO_{1-x}F_xBiCh_2$ (Ch: S, Se) [133,141]. For the system with a small RE element, the synthesis of the parent phase ($REOBiS_2$) is difficult. However, doped phases can be synthesized for RE = Sm and Yb. There is no report on the successful synthesis of $REO_{1-x}F_xBiS_2$ with other RE.

For the $REOBiS_2$ phases, single crystals were successfully grown [57,58,221]. The plate-like crystals can be easily exfoliated, which is due to the presence of a van der Waals gap between the BiS planes. The transport measurements with single crystals revealed a huge anisotropy of electronic conductivity and superconductivity ($H_{c2}$) in the $BiS_2$-based compounds.

**2.3 *SrFBiS₂ type***

$SrFBiS_2$ has a structure composed of $BiS_2$ layers and SrF BL layers as shown in Fig. 1(e) [47]. The space group of $SrFBiS_2$ is tetragonal $P4/nmm$. $SrFBiS_2$ is also a band insulator. Carrier doping is achieved by substituting $Sr^{2+}$ by $RE^{3+}$. A typical superconductor is $Sr_{1-x}La_xFBiS_2$ [51]. The basic physical and chemical properties of $SrFBiS_2$ and doped phases are similar to those of $REOBiS_2$-based systems. The Sr site can be replaced with Eu or Ca [120,182]. $EuFBiS_2$ is one of the most exotic systems because it is metallic without element substitution and shows superconductivity at 0.3 K [120]. The metallic states of stoichiometric $EuFBiS_2$ can be understood on the basis of the mixed valence state of Eu (+2.2–2.4) [120]. The S site can be substituted by Se [169,182,183,185,199]. In addition, a charge-density-wave-like (CDW-like) transition was observed in the temperature dependence of resistivity in $EuFBiS_2$ [120]. The relationship between superconductivity and CDW instability is an issue that should be clarified.

***2.4 Eu₃F₄Bi₂S₄ type***

As displayed in Fig. 1(f), $Eu_3F_4Bi_2S_4$ has a BL layer ($Eu_3F_4$ layer) thicker than the REO and SrF layers [129]. The space group of $Eu_3F_4Bi_2S_4$ is tetragonal $I4/mmm$. $Eu_3F_4Bi_2S_4$ shows superconductivity at 1.5 K. The metallic nature of $Eu_3F_4Bi_2S_4$ is understood on the basis of the mixed valence state of Eu, similarly to $EuFBiS_2$. In addition, the valence state of Eu in the $Eu_3F_4$ layer is very interesting. The valences of the inner and outer Eu sites are different. According to the different valence states, one of the Eu sites can be selectively replaced by Sr, resulting in $Eu_{3-x}Sr_xF_4Bi_2S_4$ [178,179].

***2.5 REOMS₂ type (M = Sb and In)***

Since the $BiS_2$ layer becomes a basic structure for layered superconductors, $SbS_2$- and $SbSe_2$-based compounds were also developed [213,215]. From theoretical calculations, $SbCh_2$-based compounds are expected to be a Dirac system [107] or high-performance thermoelectric materials [110]. In addition, from the similar electronic structure to $BiS_2$-based compounds, superconductivity can also be expected. As shown in Fig. 1(i), the $REOSbCh_2$ phases can be obtained with the layered structure similar to $REOBiCh_2$. However, at present, the obtained samples in polycrystalline and single-crystal forms are insulating. The insulating behavior can be related to the huge in-plane disorder, which will be explained later in detail with the $BiS_2$-based systems. To induce the conducting nature expected from theoretical studies, further material development is required according to the chemical pressure concept. The space group of $REOSbS_2$ is tetragonal $P4/nmm$ for RE = Nd and Sm and monoclinic $P2_1/m$ for La = La and Ce. For $SbSe_2$-based compounds, the space group is tetragonal $P4/nmm$. Figure 1(i) depicts the tetragonal structure.

Miura et al. synthesized $LaOInS_2$ [218]. The crystal structure is different from that of $BiCh_2$- or $SbCh_2$-based compounds. The In site was refined by the splitting model, which suggests a huge atomic displacement from the center position. The space group of $LaOInS_2$ is orthorhombic $P2_1/m$ ($\beta$ was fixed as 90º). $LaOInS_2$ is an insulator with a band gap of ~2.5 eV and shows photocatalytic activity. In addition, the $InS_2$ layer has no van der Waals gap, which is present in the $BiCh_2$- and $BiSb_2$-based compounds. The difference may be explained by the absence/presence of lone pair electrons, $5s^0$ for $In^{3+}$, $6s^2$ for $Bi^{3+}$, and $5s^2$ for $Sb^{3+}$, in these compounds.

### *2.6 $La_2O_2M_4S_6$ type*

Sun et al. synthesized the $LaOBiPbS_3$ phase containing $M_4S_6$ conducting layers (M: Pb, Bi, Ag, and other metals) and $La_2O_2$ BL layers [193]. As shown in Fig. 1(k), $LaOBiPbS_3$ is composed of LaO layers and $BiPbS_3$ layers. From synchrotron X-ray and neutron diffraction, we revealed that the $BiPbS_3$ layer is ordered into Bi-rich and Pb-rich sites. Therefore, as described in Fig. 1(k), the structure can be regarded as stacks of $LaOBiS_2$-type and rock-salt-type (PbS-type) layers [194]. The space group of $LaOBiPbS_3$ is tetragonal $P4/nmm$. From band calculations, $LaOBiPbS_3$ was predicted to be close to a zero-gap semiconductor (or metal), but the obtained sample shows insulating behavior at low temperatures.

According to the structural (stacking) concept in $LaOBiPbS_3$, we synthesized $La_2O_2Bi_3AgS_6$ by replacing the PbS layer with the (Ag,Bi)S layer [205]. As shown in Fig. 1(l), the (Ag,Bi)S-type layer is inserted between $LaOBiS_2$-type layers. Surprisingly, $La_2O_2Bi_3AgS_6$ shows metallic nature even at low temperatures and shows superconductivity at 0.5 K [214]. $La_2O_2Bi_3AgS_6$ also shows a CDW-like transition, which is similar to the case of $EuFBiS_2$.

## 3. Superconductivity in $REO_{1-x}F_xBiCh_2$

### *3.1 Electron doping*

As mentioned above, the parent phases of the $BiS_2$-based superconductors are a band insulator. Since the conduction band of $BiS_2$-based compounds is basically composed of Bi $6p_x$ and $6p_y$ orbitals hybridized with S 3p orbitals, electron doping is effective to induce a metallic nature. The typical of doping electrons in $LaOBiS_2$ is the partial substitution of $F^-$ for the $O^{2-}$ site. As shown in Fig. 3, electron carriers are generated in $BiS_2$ layers owing to the compensation of the charge neutrality. Electron carrier doping can be achieved by manipulating the BL layer of $REOBiCh_2$. By substituting $La^{3+}$ by $M^{4+}$ (M = Th, Ti, Zr, and Hf), electron carriers can be generated in the $BiS_2$ layers [52]. In addition, the mixed valence states of Ce can also be effective in generating electron carriers, which has been observed in $CeOBiS_2$ and $La_{1-x}Ce_xOBiSSe$ [171,196].

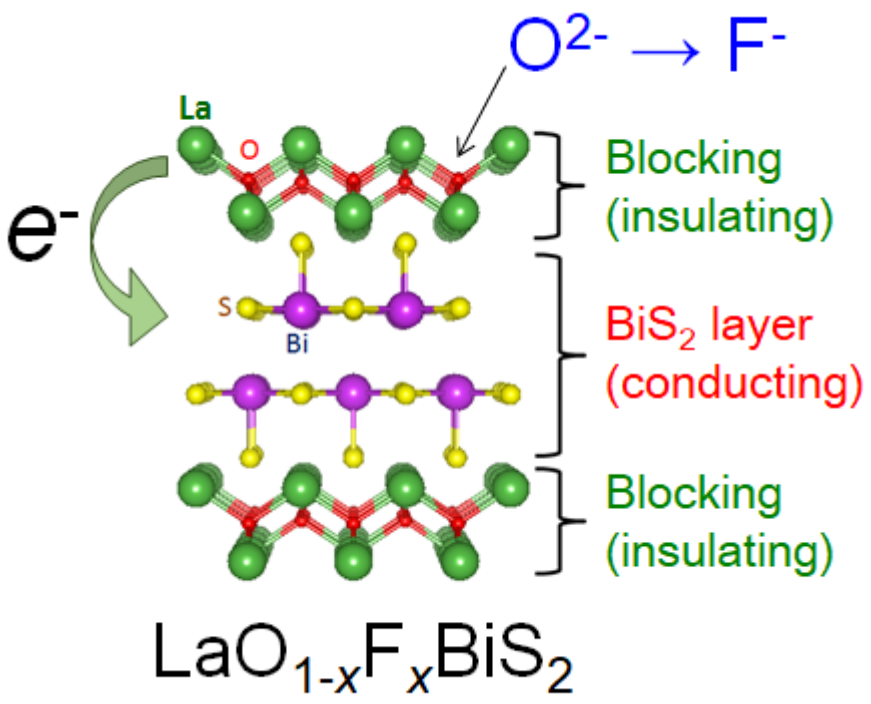

Fig. 3. (Color online) Schematic image of electron doping by F substitution in $LaO_{1-x}F_xBiS_2$.

### *3.2 Carrier doping in $LaO_{1-x}F_xBiS_2$*

Figure 4(a) shows the temperature dependence of the electrical resistivity of $LaOBiS_2$ and F-substituted $LaO_{1-x}F_xBiS_2$ [13]. $LaOBiS_2$ shows semiconducting-like behavior: the resistivity increases with decreasing temperature. Although $LaO_{0.5}F_{0.5}BiS_2$ also shows a slight increase in resistivity at low temperatures, the absolute value of resistivity for $LaO_{0.5}F_{0.5}BiS_2$ is clearly lower than that of the parent phase. In $LaO_{0.5}F_{0.5}BiS_2$, a superconducting transition is observed at 2.5 K. Although a superconducting transition is observed in the resistivity, the shielding volume fraction estimated from the magnetic susceptibility is quite small for a bulk superconductor [Fig. 4(b)]. However, we observed a slight decrease in susceptibility from below 10 K, which was the indication of the presence of local phases with a higher $T_c$. To induce bulk superconductivity in $LaO_{0.5}F_{0.5}BiS_2$, crystal structure modification by a high pressure (or chemical pressure) is required.

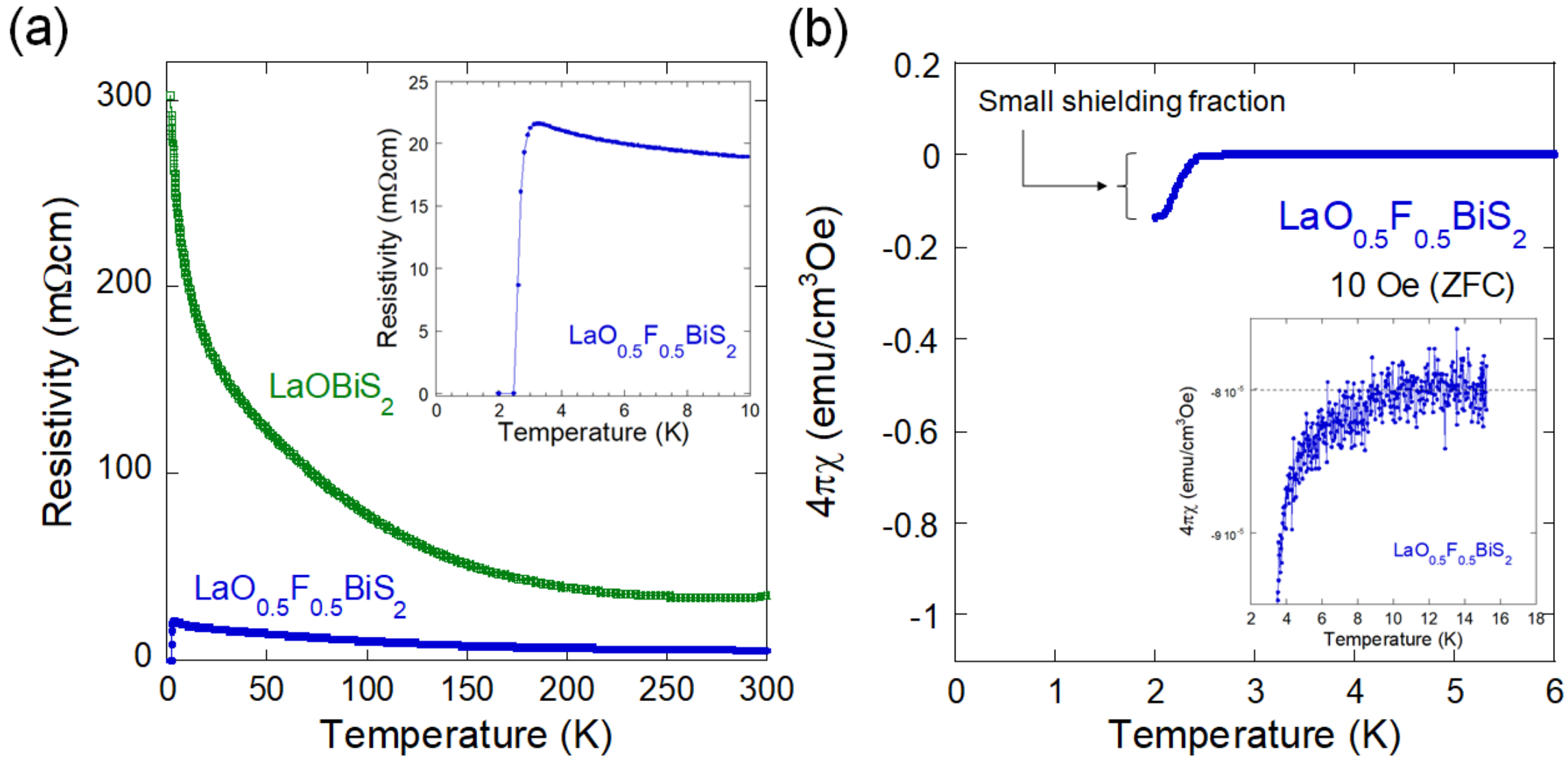

Fig. 4. (Color online) Temperature dependences of (a) electrical resistivity and (b) magnetic susceptibility of $LaOBiS_2$ and $LaO_{0.5}F_{0.5}BiS_2$. The original data was partly published in Ref. 13.

### *3.3 High-pressure annealing for $LaO_{0.5}F_{0.5}BiS_2$*

Bulk superconductivity in $LaO_{0.5}F_{0.5}BiS_2$ can be induced by annealing the as-grown powder under high pressure (HP) [13,50]. Figure 5(a) shows the temperature dependences of electrical resistivity for as-grown and HP-annealed samples of $LaO_{0.5}F_{0.5}BiS_2$. Surprisingly, $T_c$ markedly increases to 10.6 K. In addition, the shielding volume fraction is markedly enhanced [Fig. 5(b)], which indicates the emergence of bulk superconductivity. Furthermore, the $LaO_{0.5}F_{0.5}BiS_2$ sample prepared by single-step HP synthesis showed the highest $T_c$ of 11.5 K among the $BiS_2$-based compounds [82]. The enhancement of $T_c$ by HP annealing or synthesis is related to the introduction of strain in the grains. Actually, the X-ray diffraction peaks, particularly the 00*l* peaks, are broadened (with a shoulder structure in the high-angle side of the peak) after HP annealing. The disorder (strain) was analyzed by pair distribution function UPDF) analysis [140]. From the PDF analysis, the intrinsic nature of in-plane structural fluctuations and the possibility of charge fluctuations linked to the superconductivity were proposed. However, the HP-annealed samples were not homogeneous because a certain degree of the pressure (strain) effect should be released after removing the pressure (after HP synthesis). Therefore, we examined the crystal structure evolution for $LaO_{0.5}F_{0.5}BiS_2$ under high pressure.

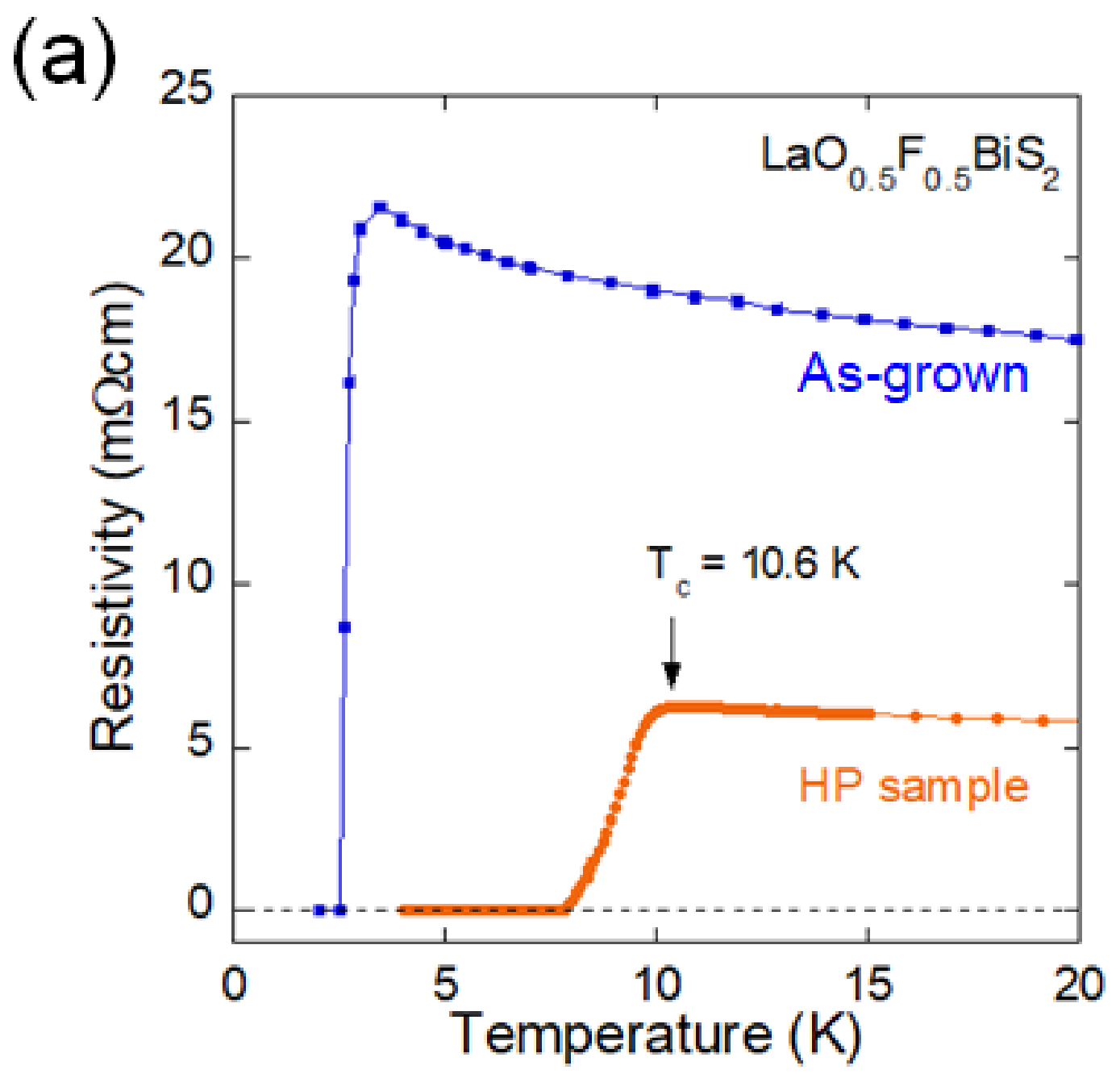

Fig. 5. (Color online) Temperature dependences of (a) electrical resistivity and (b) magnetic susceptibility for as-grown and HP-annealed samples of $LaO_{0.5}F_{0.5}BiS_2$. The original data was partly published in Ref. 13.

### *3.4 High-pressure effect*

The high-$T_c$ phase of $LaO_{0.5}F_{0.5}BiS_2$ can be obtained by applying HP to the as-grown sample [55,56,81]. Figure 6(a) shows the temperature dependences of electrical resistivity for $LaO_{0.5}F_{0.5}BiS_2$ (as-grown sample) under high pressure. The semiconducting-like behavior observed at ambient pressure is suppressed with increasing pressure. Suddenly, the high-$T_c$ phase appears at around 0.8 GPa. Figure 6(b) shows the pressure dependence of $T_c$ for $LaO_{0.5}F_{0.5}BiS_2$. From the pressure phase diagram, it is clear that there is a critical pressure, which is related to a structural transition. From the X-ray diffraction under HP, a structural transition from tetragonal (*P*4/*nmm*) to monoclinic ($P2_1/m$) was revealed [81]. A schematic image of the structural change is shown in Fig. 6(c). To discuss the difference in local structure between tetragonal and monoclinic phases, we focus on the BiS square network (BiS plane). The BiS plane slides, and the interplane Bi-Bi distance becomes shorter, as if a Bi-Bi bonds were formed. Furthermore, in the monoclinic structure, the BiS plane can be regarded as Bi-S zigzag chains. This pressure study revealed that the two-dimensional Bi-S network is not required for the emergence of superconductivity, and the quasi-one-dimensional Bi-S zigzag chain can be superconducting with a higher $T_c$. The relationship between the square network and the zigzag chains in the BiS plane will be discussed with Fig. 11 later.

The monoclinic structure should be related to the disorder (strain) induced by the HP annealing or synthesis. Basically, $BiS_2$-based compounds possess a structural instability [89], probably due to the presence of lone-pair electrons of Bi [155]. In addition, on the basis of temperature dependence measurements on local disorder for $LaOBiS_{2-x}Se_x$ [222], the structural instability can be considered as static structural disorder. This structural instability is a key concept to understand the emergence of superconductivity, which will be explained in Sect. 4.

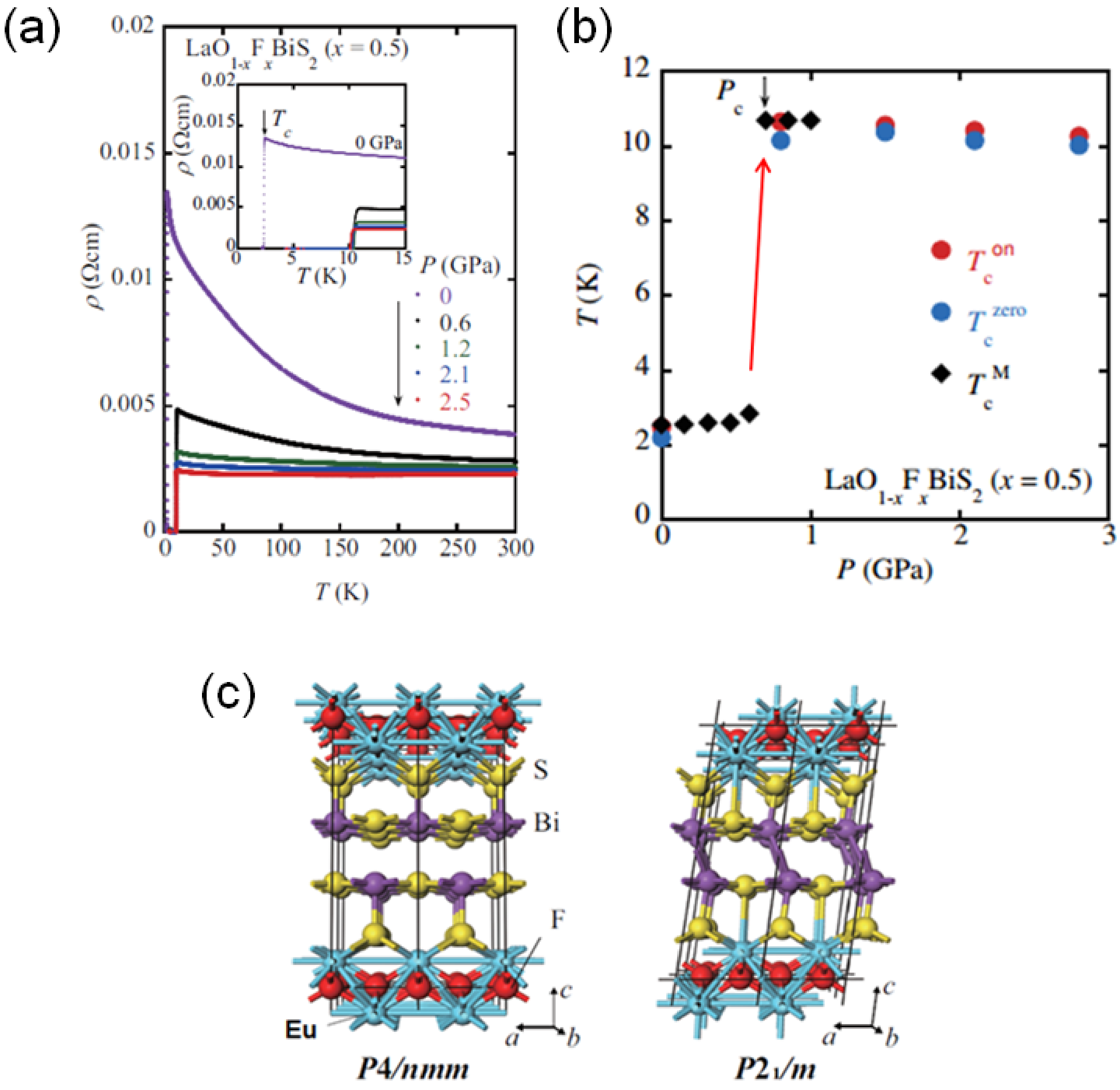

Fig. 6. (Color online) (a) Temperature dependences of electrical resistivity for $LaO_{0.5}F_{0.5}BiS_2$ under high pressure. (b) Pressure dependence of $T_c$ for $LaO_{0.5}F_{0.5}BiS_2$. (c) Schematic image of structural transition from tetragonal $P4/nmm$ to monoclinic $P2_1/m$ by pressure effect. Reproduced from Tomita et al., J. Phys. Soc. Jpn. 83, 063704 (2014). [81]. ©[2014] The Physical Society of Japan.

### *3.5 Structural transition*

The structural transition from tetragonal to monoclinic occurs in parent (non-electron-doped) phases such as $LaOBiS_2$ and Se-substituted $LaOBiS_{2-x}Se_x$. Sagayama et al. revealed that the low-temperature (below room temperature) structure of $LaOBiS_2$ was monoclinic $P2_1/m$ from single-crystal structural analysis with synchrotron X-rays [161]. From powder diffraction with synchrotron X-rays, the structural transition of $LaOBiS_{2-x}Se_x$ was also confirmed. (See Ref. 184 for $x$ = 1.0 at 300 K. Other data at compositions and temperatures are presented in this article for the first time.) Basically, the electron-doped phases are tetragonal, and the phases with a small lattice parameter ($a$) tend to be crystalized in the tetragonal structure. Owing to the intrinsic structural instability, probably due to the presence of Bi lone-pair electrons and an interlayer misfit effect, $BiS_2$-based compounds show structural flexibility with the tetragonal and monoclinic structures. This trend could be an origin of the huge HP annealing effect in $LaO_{0.5}F_{0.5}BiS_2$.

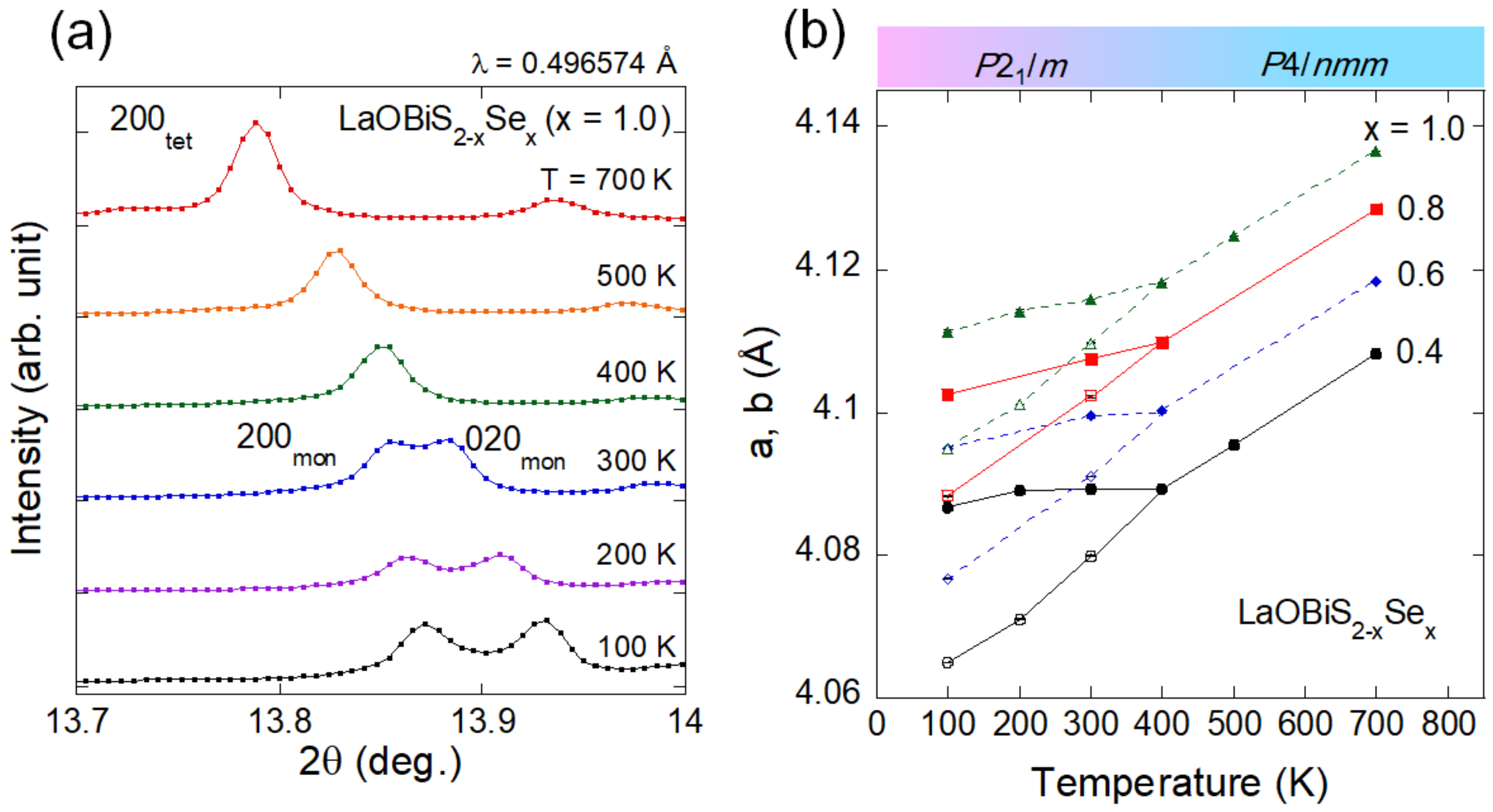

Fig. 7. (Color online) (a) Temperature dependence of synchrotron X-ray profile near the 200 peak of $LaOBiS_{2-x}Se_x$ for $x = 1.0$. Tet and mon denote tetragonal and monoclinic, respectively. (b) Temperature dependences of lattice parameters for $LaOBiS_{2-x}Se_x$.

### *3.6 Carrier doping in $NdO_{1-x}F_xBiS_2$*

As introduced above, the emergence of bulk superconductivity in $LaO_{1-x}F_xBiS_2$ requires the optimization of two parameters. One of the important parameters is electron doping because the parent phase is basically an insulator with a band gap. In addition, crystal structure optimization by the HP effects is required to induce bulk superconductivity. In contrast, $NdO_{1-x}F_xBiS_2$ exhibits a simple superconductivity phase diagram against only the carrier doping amount. A comparison of the phase diagram between $LaO_{1-x}F_xBiS_2$ and $NdO_{1-x}F_xBiS_2$ is shown in Fig. 8. Although filamentary superconductivity is observed for $LaO_{1-x}F_xBiS_2$ (as-grown), bulk superconductivity is observed in a wide range of carrier doping ($x$) for $NdO_{1-x}F_xBiS_2$. The difference between these two phases is the in-plane lattice parameter ($a$-axis). Since the ionic radius of $Nd^{3+}$ (112 pm) is smaller than that of $La^{3+}$ (116 pm), assuming a coordination number of 8, the $BiS_2$ layer is compressed in $NdO_{1-x}F_xBiS_2$ owing to the shrinkage of the BL layer from La(O,F) to Nd(O,F). This is a kind of chemical pressure effect because both $La^{3+}$ and $Nd^{3+}$ are trivalent in this crystal structure; hence, the carrier doping amount is controlled by adjusting the F concentration ($x$) only. Motivated by this chemical pressure effect, we systematically investigated two systems with a tunable chemical pressure amplitude.

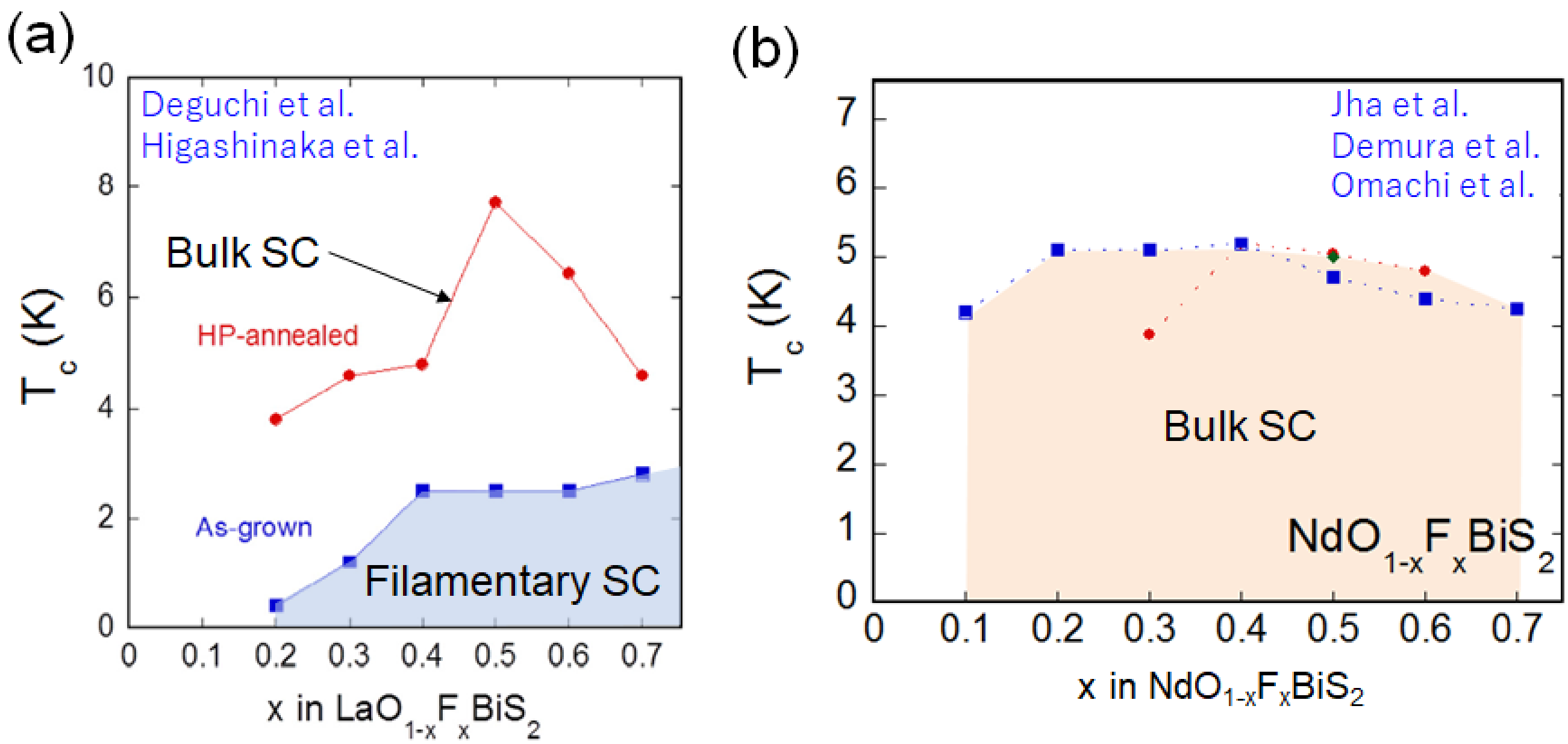

Fig. 8. (Color online) Superconductivity phase diagram for (a) $LaO_{1-x}F_xBiS_2$ and (b) $NdO_{1-x}F_xBiS_2$. The original data were partly published in Refs. 17, 46, 50, 85, and 138. SC denotes superconductivity.

### *3.7 Chemical pressure effect*

Chemical pressure (CP) is used for the lattice volume manipulation of materials using isovalent (partial or full-site) substitution. Using the CP effect, we can obtain effects similar to those obtained using the HP technique. The merit of the CP effect is the great flexibility and precision of lattice volume tuning. In addition, we can avoid undesired disorder (or strain) effects, which sometimes occur in HP synthesis [13,140]. For the $REOBiCh_2$ systems, two typical CP effects are powerful for systematically changing the physical properties.

As mentioned earlier, RE-site substitution with $RE^{3+}$ with a different ionic radius [Fig. 9(a)] is a powerful method. The superconductivity phase diagram is shown in Fig. 9(c). Although the carrier doping (nominal $x$) is fixed as $x$ = 0.5 (0.5 electron per Bi), the superconducting property markedly changes with decreasing RE ionic radius from La to Sm [127,141]. Notably, the filamentary superconducting phase for RE = La (as-grown) is suppressed at RE = $La_{0.5}Ce_{0.5}$, and bulk superconductivity emerges at RE = $Ce_{0.6}Nd_{0.4}$. $T_c$ increases with increasing CP.

The other CP effect is Se substitution for the in-plane S site [Fig. 9(b)]. In the $REOBiCh_2$ structure, the S site can be systematically substituted by Se, and the Se ion preferentially occupies the in-plane Ch1 site [133,141]. As shown in Fig. 9(d), bulk superconductivity is induced by the CP effect with Se substitution in $LaO_{0.5}F_{0.5}BiS_{2-x}Se_x$ [128,141]. Interestingly, in the low-Se region, $T_c$ decreases at $x$ = 0.2 (the shielding fraction also decreases here), and bulk superconductivity is induced with further Se substitution. This trend is clearly similar to that observed in $REO_{0.5}F_{0.5}BiS_2$ [Fig. 9(c)]. From the commonality in the CP phase diagrams, the link between CP, superconductivity, and local structure is suggested. To clarify the local structure changes and thier relation to the emergence of bulk superconductivity in this system, we performed synchrotron X-ray diffraction and X-ray absorption spectroscopy (XAS) for these CP samples.

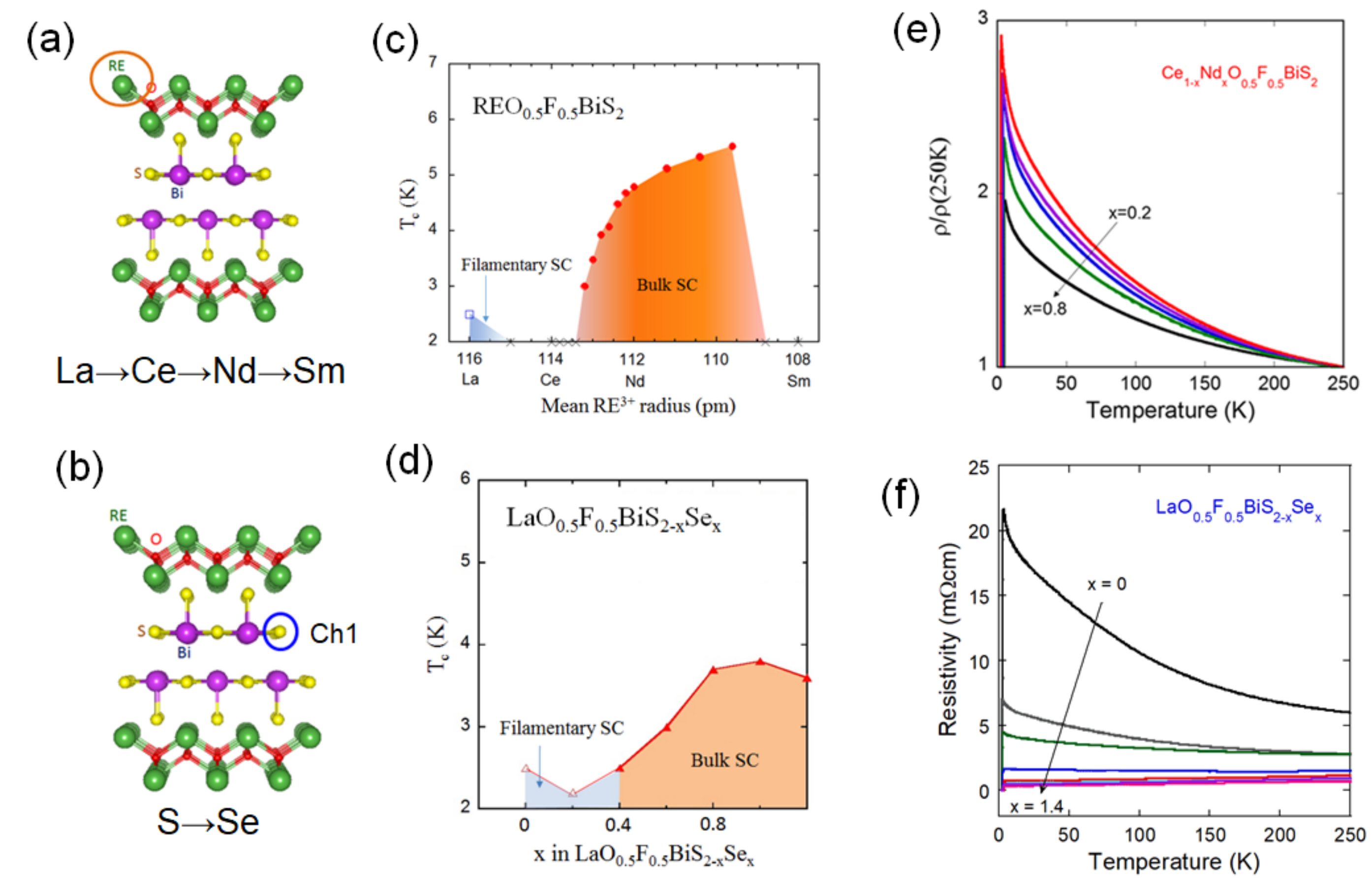

Fig. 9. (Color online) (a) Schematic image of RE site substitution. (b) Schematic image of Se substitution for the in-plane Ch1 site. (c,d) Superconductivity phase diagram for $REO_{0.5}F_{0.5}BiS_2$ and $LaO_{0.5}F_{0.5}BiS_{2-x}Se_x$. SC denotes superconductivity. (e,f) Temperature dependences of normalized resistivity for $REO_{0.5}F_{0.5}BiS_2$ and $LaO_{0.5}F_{0.5}BiS_{2-x}Se_x$.

## 4. Importance of Local Structure

### *4.1 In-plane chemical pressure*

From synchrotron X-ray diffraction and Rietveld refinement, we discuss the correlation between crystal structure parameters and superconducting properties [see phase diagrams in Figs. 9(c) and 9(d)]. Since these phase diagrams exhibit similar trends as explained above, there should be common structural features in their CP effects. However, although we expected that the in-plane bond distance or the Ch1-Bi-Ch1 angle (flatness of the BiCh1 plane) is essential for the emergence of superconductivity, there was no clear correlation. As shown in Figs. 10(a) and 10(b), the substitution of small RE in $REO_{0.5}F_{0.5}BiS_2$ results in lattice shrinkage. In contrast, the Se substitution in $LaO_{0.5}F_{0.5}BiS_{2-x}Se_x$ results in lattice expansion because the ionic radius of $Se^{2-}$ (198 pm) is larger than that of $S^{2-}$ (184 pm). However, the phase diagrams are similar.

This contradiction can be solved by introducing the concept of in-plane CP amplitude. We estimated the in-plane bond length (Bi-Ch1 distance) and calculated the total of the Bi ionic radius and the Ch1 ionic radius (average). The in-plane CP amplitude parameter was defined as CP = (Bi-Ch1 distance) / ($R_{Bi}$ + $R_{Ch}$), where $R_{Bi}$ and $R_{Ch}$ are the ionic radii of Bi and Ch in the Bi-Ch1 plane, respectively. (Please see original paper discussing the in-plane CP effect [141].) The estimated in-plane CP amplitude is plotted in Fig. 10(d). Interestingly, in-plane CP increases by substitution for both systems. In Fig. 10(d), bulk superconductivity phases are indicated by a yellow region. Therefore, bulk superconductivity is induced at around the CP = 1.01 for both systems. Although the values of in-plane CP may not be directly related to the physical parameters, the CP is useful for the qualitative understanding of the emergence of bulk superconductivity in $REO_{0.5}F_{0.5}BiCh_2$. In Fig. 10(c), a schematic image of the in-plane CP effect is displayed. For $REO_{0.5}F_{0.5}BiS_2$, the $BiS_2$ layer is compressed with decreasing *a*-axis length by a smaller Nd substitution. Since the composition of the SC layer is not modified, the shrinkage of the Bi-S1 distance directly enhances the in-plane CP amplitude. For $LaO_{0.5}F_{0.5}BiS_{2-x}Se_x$, the composition of the La(O,F) layer does not change, but the Se concentration in the Bi-Ch1 plane is modified according to $x$. Basically, the bonding of the La(O,F) layer is stronger than that of the flexible $BiCh_2$ layer. Therefore, the *a*-axis cannot expand significantly, even though the larger $Se^{2-}$ ion filled the space of the two-dimensional network. This misfit strain effect generates a strong in-plane CP effect in $LaO_{0.5}F_{0.5}BiS_{2-x}Se_x$.

$T_c$ also shows a good correlation with the in-plane CP amplitude as shown in Fig. 11(a). With increasing CP, $T_c$ increases and approaches 6 K. However, the phase with RE = Sm is not a superconductor [126]. The estimated CP is close to that for RE = Ce, which is also non-superconducting. This should be due to the mixed valence states of Sm and a larger ionic radius.

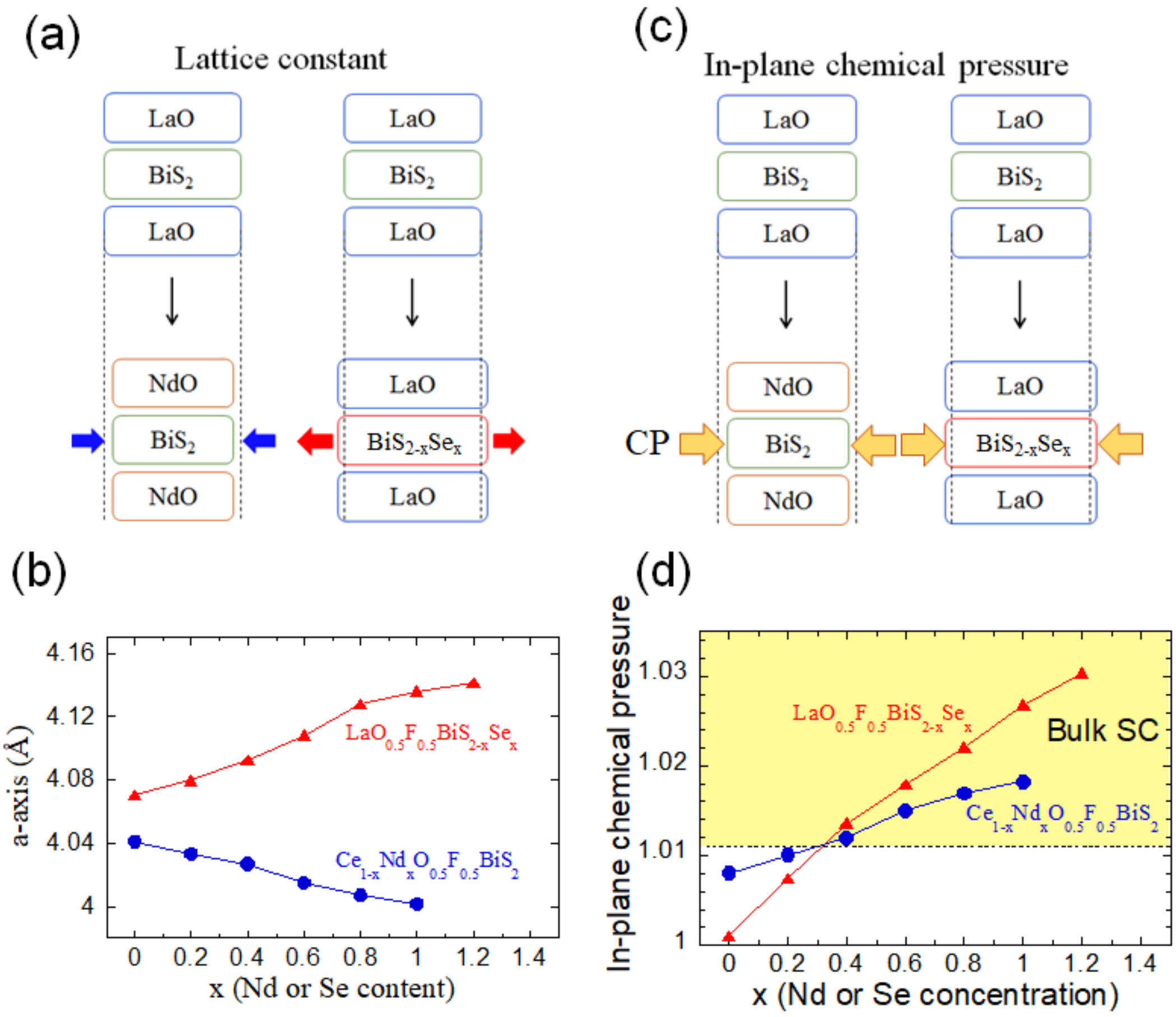

Fig. 10. (Color online) (a) Schematic image of lattice shrinkage and expansion in $REO_{0.5}F_{0.5}BiS_2$ and $LaO_{0.5}F_{0.5}BiS_{2-x}Se_x$. (b) Lattice parameter (*a*-axis) for $REO_{0.5}F_{0.5}BiS_2$ and $LaO_{0.5}F_{0.5}BiS_{2-x}Se_x$. (c) Schematic image of in-plane CP effect in $REO_{0.5}F_{0.5}BiS_2$ and $LaO_{0.5}F_{0.5}BiS_{2-x}Se_x$. (d) In-plane CP amplitude for for $REO_{0.5}F_{0.5}BiS_2$ and $LaO_{0.5}F_{0.5}BiS_{2-x}Se_x$. SC denotes superconductivity, and the yellow colored region indicates a bulk SC phase.

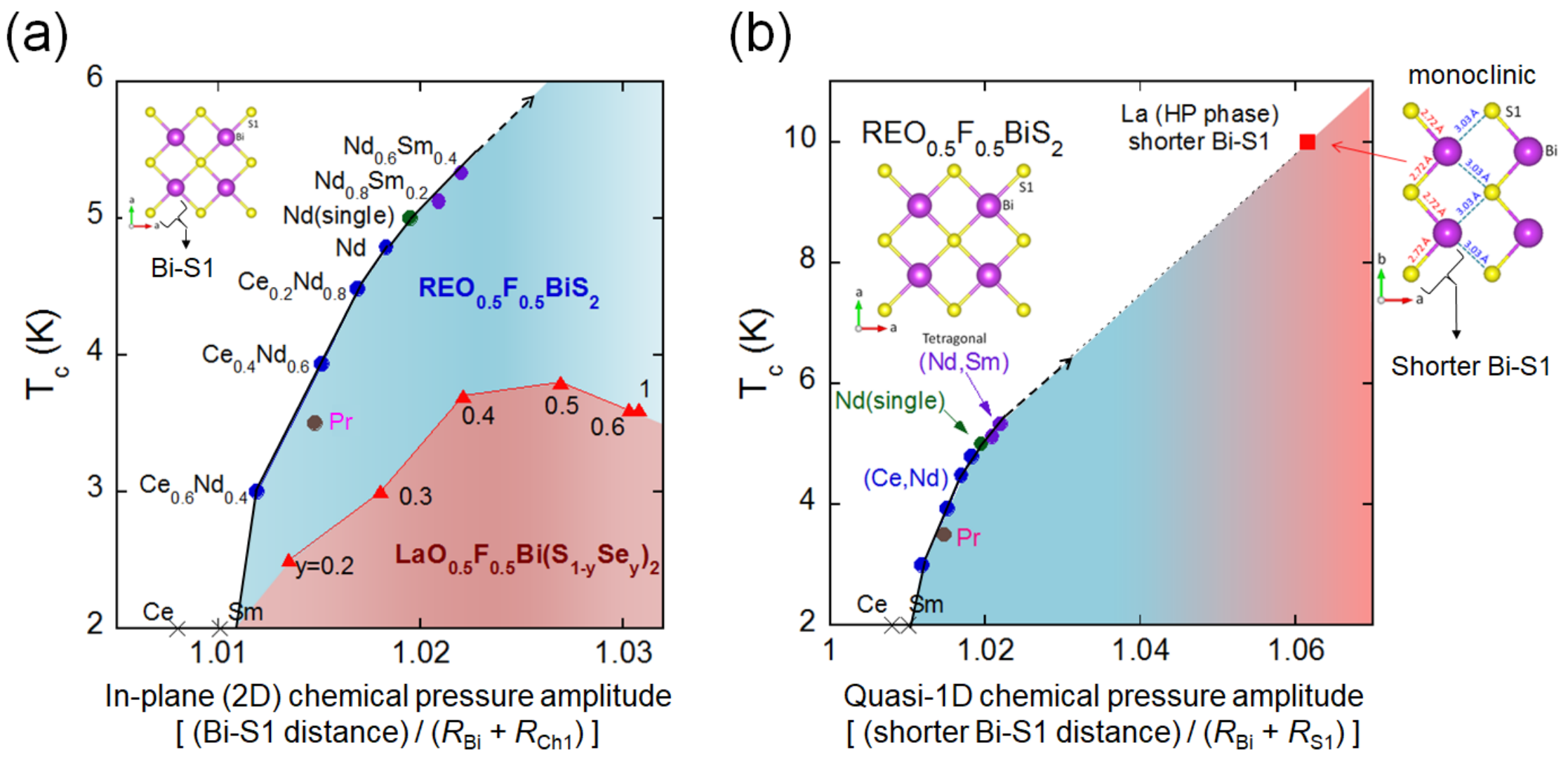

Fig. 11. (Color online) Superconductivity phase diagrams for $REO_{0.5}F_{0.5}BiS_2$ and $LaO_{0.5}F_{0.5}BiS_{2-x}Se_x$ as functions of (a) in-plane (2D) CP amplitude (estimated using the equation described at the bottom) and (b) quasi-one-dimensional CP amplitude. The original data was partly published in Ref. 141. 2D and 1D denote two- and one-dimensional, respectively.

To relate the CP effects to the HP effects, the data for the HP phase of $LaO_{0.5}F_{0.5}BiS_2$ is added into the phase diagram. The HP phase has a zigzag chain, and there are two different Bi-Ch1 distances. The shorter distance is the in-chain bond of 2.72 Å, and the longer distance is the interchain bond of 3.03 Å. Assuming that the shorter bond is essential for the superconductivity, we plotted the quasi-one-dimensional CP phase diagram. Although there is only one data point for the HP phase, it seems to be located at the extension of the two-dimensional plots. A two-dimensional Bi-Ch network is not essentially for the emergence of superconductivity in the $BiCh_2$-based compounds. To increase $T_c$, a shorter Bi-Ch1 distance is required.

*4.2 Suppression of in-plane disorder at the Ch1 site*

So far, the importance of both carrier doping and the in-plane (or quasi-one-dimensional) CP effect has been introduced. However, the structural parameter directly linked to the emergence of bulk superconductivity should be identified. This was achieved by investigating the anisotropic atomic displacement parameters. For the anisotropic analysis of the displacement parameter $U$, the in-plane $U_{11}$ and the out-of-plane $U_{33}$ are refined by Rietveld refinement. As shown in Fig. 12(a), $U_{11}$ for S1 for $LaO_{0.5}F_{0.5}BiS_2$ is clearly larger than $U_{11}$ for Bi [206]. With decreasing RE ionic radius, $U_{11}$ for S1 decreases, and the bulk superconductivity is induced when the in-plane S1 disorder is suppressed. Similarly, Se substitution can suppress the in-plane disorder of the Ch1 site in $LaO_{0.5}F_{0.5}BiS_{2-x}Se_x$ [184]. In addition, from XAS, the suppression of in-plane Bi-Ch1 disorder with increasing Se was detected [186]. The correlation between the CP effect and $U_{11}$ for Ch1 can be understood from Fig. 12(c), which shows the carrier doping dependences of $U_{11}$ for $LaO_{1-x}F_xBiSSe$. In $LaO_{1-x}F_xBiSSe$, the composition of the BiSSe layer is not modified, but the F concentration was systematically changed to increase the carrier doping. Since the BiSSe layer possesses a high amplitude of in-plane CP, the $U_{11}$ values for Ch1 and Bi are small and comparable for all $x$. Namely, the carrier doping is not linked to the in-plane disorder. The evolutions of anisotropic displacement parameters are shown in Fig. 13. The in-plane $U_{11}$ for the Ch1 site is clearly suppressed by the CP effects for $REO_{0.5}F_{0.5}BiS_2$ and $LaO_{0.5}F_{0.5}BiS_{2-x}Se_x$ (with the crystal structures of $NdO_{0.5}F_{0.5}BiS_2$ and $LaO_{0.5}F_{0.5}BiSSe$).

Using the $LaO_{1-x}F_xBiSSe$ system with less in-plane disorder, we can investigate the essential phase diagram for carrier doping and superconductivity for the $BiCh_2$-based systems. Figure 12(d) shows the superconductivity phase diagram for $LaO_{1-x}F_xBiSSe$ [184]. With a small amount of electron carriers, metallic conductivity and bulk superconductivity are induced. The evolution of $T_c$ shows an anomalous flat dependence on carrier concentration. This is consistent with the presence of correlation between $T_c$ and the in-plane Bi-Ch1 bond distance. These trends suggest that the optimization of the crystal structure is very important for increasing $T_c$ in the $BiCh_2$-based superconductors.

Let us comment on the intrinsic structural disorder in $BiS_2$-based compounds. As mentioned above, the intrinsic disorder is likely to be caused by the presence of Bi lone-pair electrons. This has been also observed for single crystals. Owing to the intrinsic inhomogeneity and local breaking of symmetry, no NMR results on the superconductivity of this system have been reported so far.

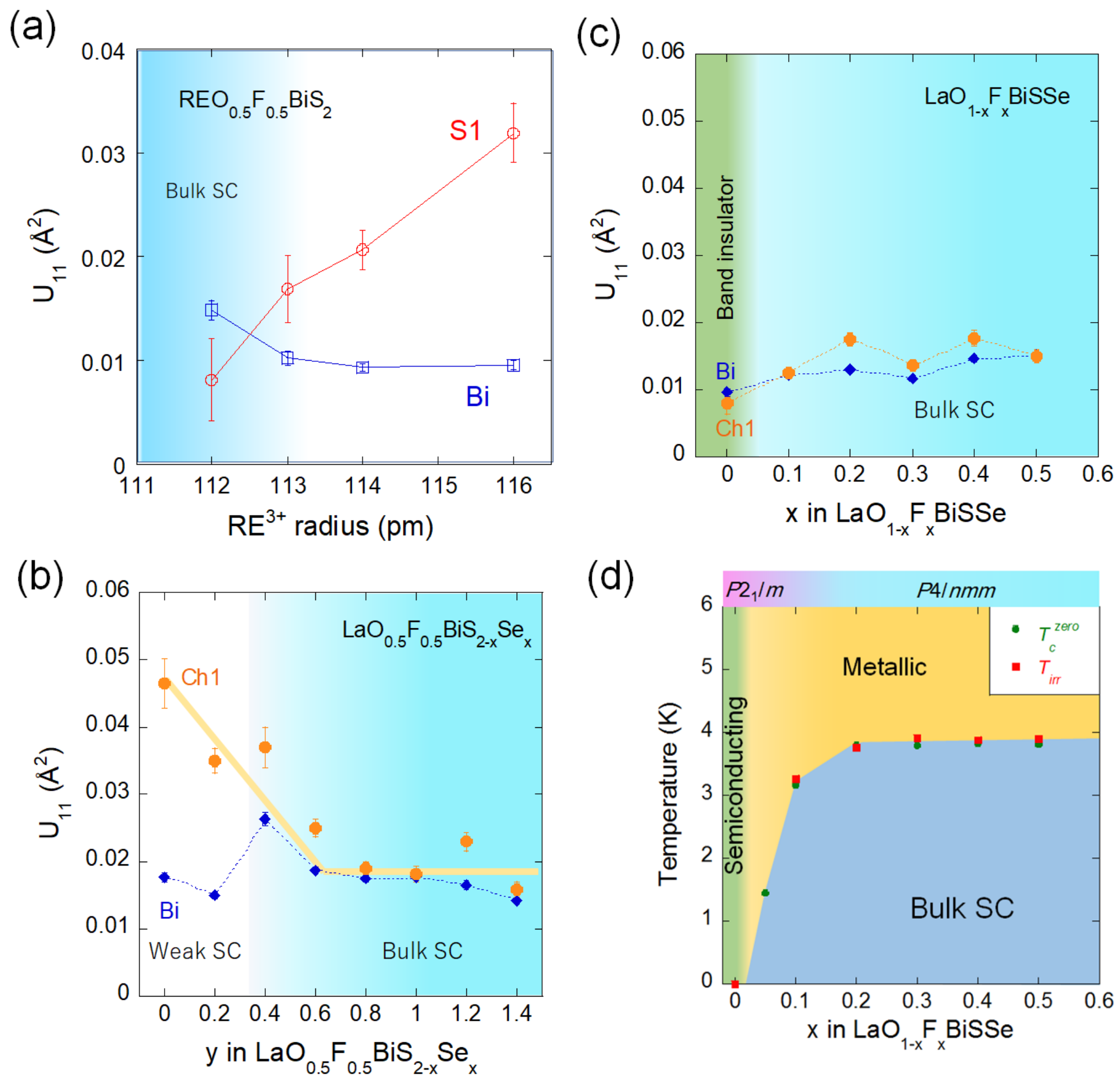

Fig. 12. (Color online) (a) RE radius dependences of $U_{11}$ (S1 and Bi sites) for $REO_{0.5}F_{0.5}BiS_2$. (b) Se concentration dependences of $U_{11}$ (Ch1 and Bi sites) for $LaO_{0.5}F_{0.5}BiS_{2-x}Se_x$. (c) Carrier doping dependences of $U_{11}$ (Ch1 and Bi sites) for $LaO_{1-x}F_xBiSSe$. The $U_{11}$ values plotted in (a)–(c) were obtained from the Rietveld analysis of synchrotron X-ray diffraction data at 300 K. (d) Essential phase diagram of carrier doping and superconductivity in $LaO_{1-x}F_xBiSSe$. The original data was partly published in Refs. 184 and 206. ©[2017] The Physical Society of Japan.

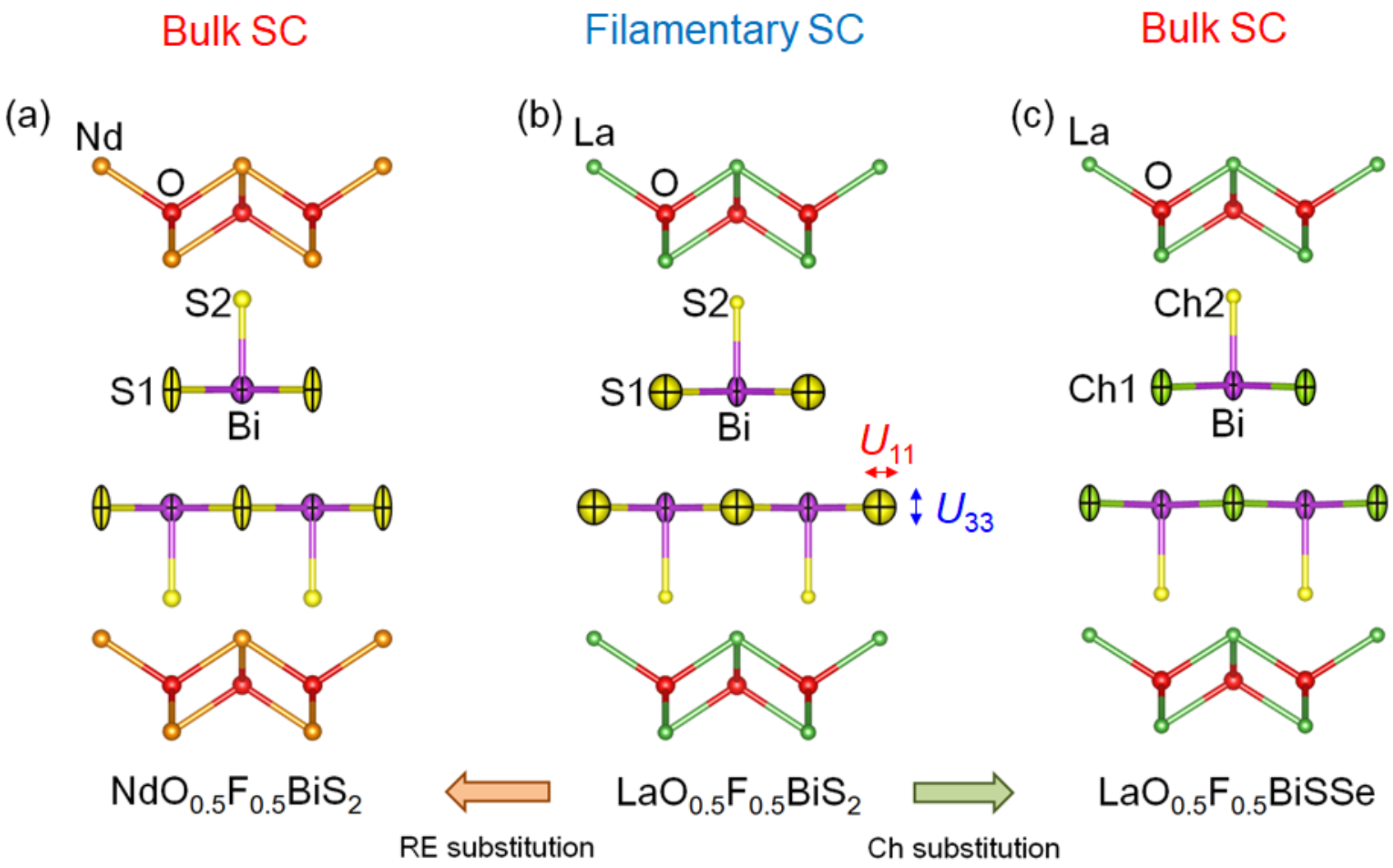

Fig. 13. (Color online) Schematic images of CP effect and anisotropic displacement parameters. The displacement ellipsoids depicte the 90% probability. The images were drawn using VESTA.

### *4.3 Possible relationship between atomic displacement of Bi and superconductivity*

Let us briefly introduce the possibility of a positive correlation between superconductivity and disorder at the Bi site. From the pair density function analysis of neutron and X-ray data, Athauda et al. proposed the importance of charge fluctuation for the superconductivity in $REO_{1-x}F_xBiS_2$ [164,192]. Although the in-plane disorder of S1 is negatively linked to the superconductivity, that for the Bi site may be positively linked to the superconductivity. Powder X-ray diffraction and the analysis of the anisotropic displacements also suggest that there is a correlation between the Bi displacement and $T_c$ [184,206]. Furthermore, the one-dimensional motion of Bi was revealed for the thermoelectric material $LaOBiS_{2-x}Se_x$ from neutron inelastic scattering [207]. This trend can be found in Fig. 13. $U_{33}$ is still large for Ch1 and Bi sites when $U_{11}$ is suppressed by the CP effects. The anharmonic motion of Bi may be related to the superconductivity in the $BiCh_2$-based system.

## 5. Studies Focusing on Pairing Mechanisms

### *5.1 Theoretical studies*

The mechanisms of superconductivity in the $BiS_2$-based superconductors have been extensively studied. Both conventional electron-phonon mechanisms [19,98] and unconventional mechanisms mediated by spin fluctuations, electron correlation, and spin-orbit coupling [91-93,101] have been proposed. Furthermore, recent theoretical calculations of $T_c$ suggested that conventional mechanisms mediated by phonons can achieve $T_c$ higher than 0.5 K for the $BiS_2$-based superconductors [104].

### *5.2 Spectroscopy*

Since single crystals are available, particularly for $REO_{1-x}F_xBiS_2$, angle-resolved photoemission spectroscopy (ARPES) studies have been performed [71,134,143,187,209]. On the actual Fermi surface (electron doping amount) in the crystals, good agreement between the theoretical calculation and ARPES studies was observed for RE = La [134]. However, for RE = Ce and Nd, the electron doping amount estimated from ARPES was much smaller than that expected from the nominal F concentration and theoretical calculations [71,143]. Ota et al. observed the node of the superconducting gap in $NdO_{0.71}F_{0.29}BiS_2$ [187] and proposed unconventional pairing mechanisms.

Using single crystals, scanning tunneling microscopy/spectroscopy (STM/STS) has been performed to analyze the surface structure and the electronic states of $BiS_2$-based compounds [58,75,158,200]. Machida et al. observed checkerboard-stripe electronic states for $NdO_{0.7}F_{0.3}BiS_2$ [75]. On the surface, a semiconducting gap of about 20 meV was observed. Note that the surface is composed of single $BiS_2$ layer only because the crystal is basically cleaved at the van der Waals gap between two $BiS_2$ layers. Therefore, the semiconducting states may be the essential character of the surface of $NdO_{0.7}F_{0.3}BiS_2$ with a single $BiS_2$ layer. These observations should be related to the CDW instability or other charge-ordering states. Liu et al. successfully observed a superconducting gap with STS for $NdO_{1-x}F_xBi_{1-y}S_2$ single crystals [58]. They also observed the opening of a semiconducting gap after cleaving. By removing surfaces using Ar mining, the superconducting gap was successfully observed. Notably, giant superconducting fluctuations were observed; the superconducting gap feature was observed even well above the bulk $T_c$.

### *5.3 Superfluid density*

To discuss the pairing symmetry, investigation of the temperature dependence of the superfluid density in the superconducting states is useful. By muon-spin spectroscopy measurements (μSR), full-gap states were revealed for $LaO_{0.5}F_{0.5}BiS_2$ (HP-annealed) [63]. From the fitting of the temperature dependence of the superfluid density, the anisotropic s-wave was proposed. s-wave superconductivity was proposed from magnetic penetration depth measurements and also from the thermal conductivity measurement [30,163]. The contradiction between the nodal gap observed in ARPES and the fully gapped states may be due to a disorder-induced topological change of the SC gap structure from a nodal s-wave to a nodeless s-wave [187]. In fact, the presence of defects in the BiS plane was proposed from STM. In addition, the $BiS_2$-based system is an intrinsically disordered system as explained earlier. To conclude the universal pairing mechanisms in the $BiS_2$-based superconductors, further measurements with various probes and integrated discussion are required.

### *5.4 Isotope effect*

To discuss the importance of the electron-phonon interaction in the pairing mechanisms, the isotope effect is one of the powerful methods. If the pairing mechanisms are dominated by the electron-phonon interaction, $T_c$ decreases with increasing mass of the isotope element. We used the $LaO_{0.6}F_{0.4}BiSSe$ phase to test the isotope effect with $^{76}Se$ and $^{80}Se$ isotope chemicals. As explained above, the BiSSe layer can be regarded as less disordered, and essential superconducting properties can be discussed. In addition, superconducting transitions in resistivity and susceptibility are quite sharp for this composition. As a result, no isotope effect on $T_c$ was observed for $LaO_{0.6}F_{0.4}BiSSe$ with $^{76}Se$ and $^{80}Se$ [195]. The result suggests that phonons are not essential for the pairing in $BiCh_2$-based superconductors.

## 6. Thermoelectric Properties

Thermoelectric materials can directly convert waste heat to electricity and are considered as one of the promising technologies for solving energy problems. $BiS_2$-based compounds show a relatively high performance (dimensionless figure-of-merit $ZT = S^2T/\kappa\rho$, where $S$, $T$, $\kappa$, and $\rho$ are the Seebeck coefficient, absolute temperature, thermal conductivity, and electrical resistivity, respectively). The power factor ($PF$) is given by $PF = S^2/\rho$, which is a parameter for the thermoelectric power amplitude for the compound. For $LaOBiS_2$, $PF$ markedly decreases with carrier doping owing to the decrease in $S$ [84]. However, $PF$ can be enhanced by Se substitution in $LaOBiS_{2-x}Se_x$ as shown in Fig. 14(a) [123,156,162]. This enhancement of $PF$ can be understood with the concept of in-plane CP. With increasing CP, resistivity decreases [Fig. 14(b)] owing to the suppression of local disorder like in the case of superconductors. In addition, Se substitution does not decrease $S^2$ because carrier concentration, which is directly linked to $S$ in this system.

In addition, Se substitution supresses $\kappa$. $LaOBiS_{2-x}Se_x$ with $x$ = 1.0 shows an extremely low $\kappa$ of about 1 W/mK. This $\kappa$ value is lower than that of quartz glass or the typical thermoelectric compound $Bi_2Te_3$. $\kappa$ is the sum of lattice thermal conductivity ($\kappa_{lattice}$) and carrier thermal conductivity. As shown in Fig. 14(c), Se substitution clearly suppresses $\kappa_{lattice}$. This is not caused by the randomness effect but by the enhancement of the anharmonicity by Se substitution. Figure 14(d) shows the temperature dependences of $U_{11}$ and $U_{33}$ for Bi [166]. $U_{33}$ is clearly higher than $U_{11}$. This anisotropic displacement is due to the presence of the rattling-like motion of Bi along the *c*-axis. Lee et al. reported the relationship between the phonon energy for Bi and the lattice thermal conductivity from neutron inelastic scattering study [207].

Ochi et al. theoretically predicted the possibility of a high *ZT* exceeding 2 for $SbSe_2$-based or $AsSe_2$-based systems [110]. As described in Sect. 2, $REOSbSe_2$ was synthesized by Goto et al., but the obtained samples show insulating behavior. On the basis of the CP concept, we need to suppress the structural (disorder) problems to enhance *PF*. Since we have observed extremely low $\kappa$ for $REOSbSe_2$, the increase in *PF* should achieve an extremely high *ZT*.

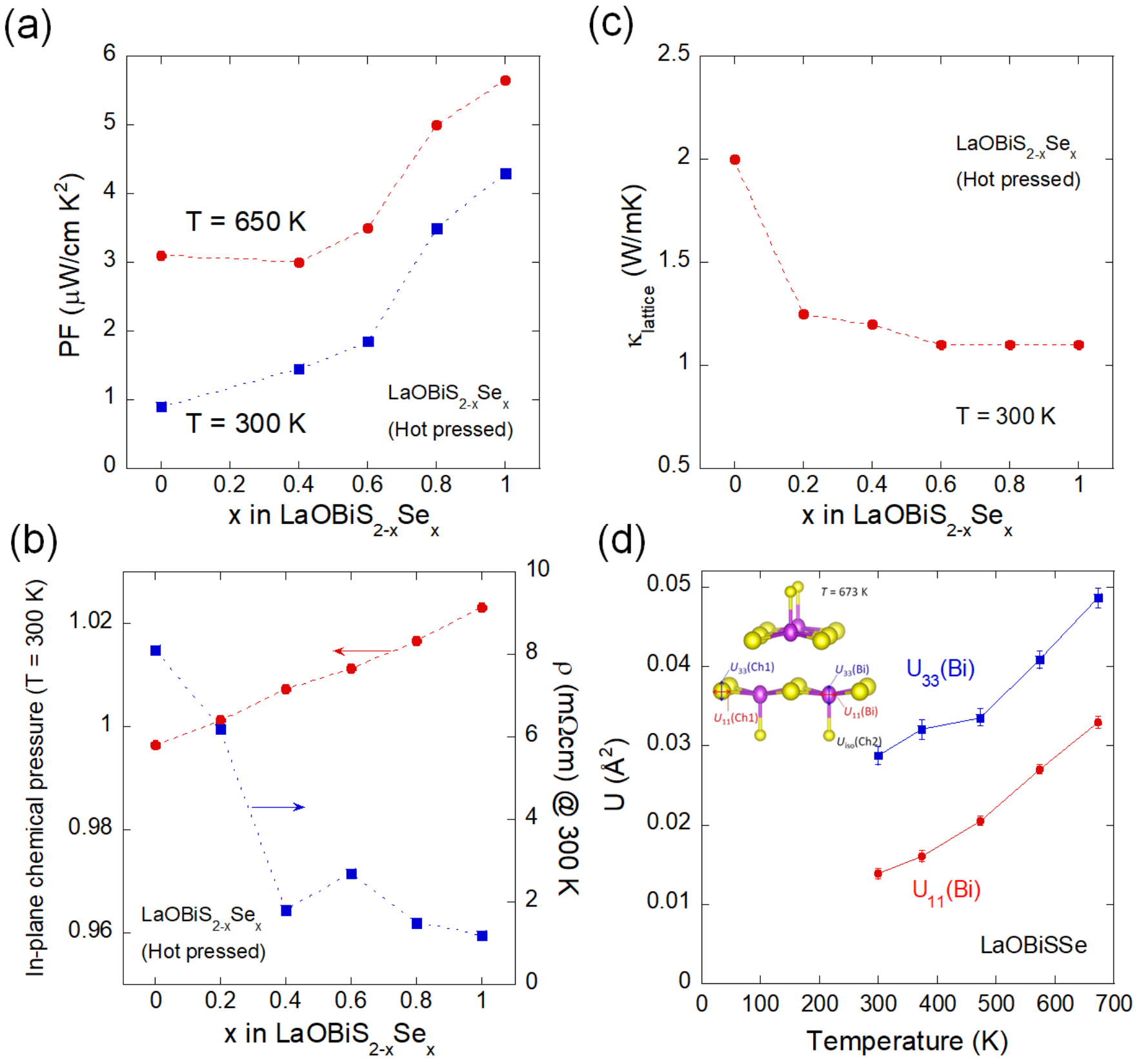

Fig. 14. (Color online) (a–c) Se concentration dependences of *PF*, in-plane CP, $\rho$, and $\kappa_{lattice}$ for $LaOBiS_{2-x}Se_x$. (d) Temperature dependences of $U_{11}$ and $U_{33}$ for Bi. The original data was partly published in Refs. 162, 166, and 207.

## 7. Summary

Since the discovery of superconductivity in the $BiS_2$-based layered compounds $Bi_4O_4S_3$ and $LaO_{1-x}F_xBiS_2$, much efforts have been made toward the material development and clarification of the mechanisms of superconductivity. In addition, for the $BiS_2$-related compounds, the functional properties of thermoelectric, spin, optical, and photocatalytic materials have also been investigated in related compounds. As summarized in Fig. 1 and Table I, various types of crystal structure and compositions have been investigated.

The crystal structure of the $BiS_2$-based compounds is composed of alternately stacked $BiS_2$ conducting (superconducting) layers and blocking (insulating) layers such as LaO or EuF. The stacking structure is similar to those of the cuprate and Fe-based superconductors. A low-dimensional electronic structure is realized, and the electronic and structural properties can be easily manipulated by element substitution and/or pressure effects, which is the merit of layered systems.

The parent phase of $BiS_2$-based superconductors is an insulator with a band gap. Since the conduction band is composed of Bi 6p orbitals hybridized with S 3p, electron doping is required to obtain a metallic character. For example, electron-doped $LaO_{1-x}F_xBiS_2$ shows superconductivity, but the superconducting states for the as-grown sample of $LaO_{1-x}F_xBiS_2$ are not bulk in nature. To induce bulk superconductivity, the optimization of the local crystal structure is important. High pressure can stabilize the monoclinic structure of $LaO_{1-x}F_xBiS_2$ and induces bulk superconductivity. In addition, the chemical pressure effect by RE substitution or Se substitution can also induce bulk superconductivity in $REO_{1-x}F_xBiCh_2$. The in-plane chemical pressure is universally essential for the emergence of bulk superconductivity in the $BiS_2$-based systems. The in-plane chemical pressure suppresses the in-plane disorder of S atoms, which intrinsically exists in the $BiS_2$-based compounds. Now, we know a method of inducing bulk superconductivity by carrier doping and structural optimization. Therefore, the next stage is the complete understanding of the mechanisms of superconductivity in the $BiS_2$-based family. There are still contradictory results on the mechanisms. Further measurements on well-designed compositions with various probes and integrated discussion are required. Finally, let us mention the possibility of a higher $T_c$ in the $BiS_2$-based systems. As shown in Fig. 4(b), small signals with a higher $T_c$ (higher than the bulk $T_c$) have been observed in some compounds. In addition, as reported by Liu et al., giant superconducting fluctuations were observed from STS. Therefore, further material development is required. We believe that a high $T_c$ can be obtained by realizing the optimized condition (possibly extremely compressed local Bi-S bonding).

**Acknowledgments**

The author would like to thank all the collaborators and researchers studying $BiS_2$-based and related compounds for their encouragement to study this new field. In particular, the author thanks his former and current group members, H. Izawa, J. Kajitani, A. Omachi, T. Hiroi, K. Nagasaka, A. Nishida, Y. Hijikata, R. Sogabe, K. Hoshi, R. Jha, Y. Goto, and Prof. O. Miura for their support; their unpublished data are partly shown in this review paper. The main results on $BiS_2$-based systems presented in this review article were obtained under the supports of Grant-in-Aid for Scientific Research (KAKENHI) (Nos. 23860042, 25707031, 26600077, 15H05886, 16H04493, and 17K19058), JST-CREST (No. JPMJCR16Q6), and the bilateral agreement between Tokyo Metropolitan University and Sapienza University of Rome. The crystal structure analysis of $LaOBiS_{2-x}Se_x$ was performed at BL02B2, SPring-8 (proposal no.: 2016B1078).

**Note**

This review article is part of a special issue on $BiS_2$-related compounds and aimed to be useful as an introductory review for the other review articles in the same issue. We hope that readers find interesting characteristics and recent advances of this system from the other review articles with specialized topics. Since the discovery of the superconductivity of $BiS_2$-based superconductors, we have continued to upload the reference information of articles that appeared in arXiv and journals to the website “Papers on $BiS_2$-based superconductors (http://www.comp.tmu.ac.jp/eeesuper/BiS2_papers.html)”. We will be happy if this database is useful for your future works.

*E-mail: mizugu@tmu.ac.jp

**References**

1. J. G. Bednorz and K. A. Müller, Z. Phys. B-Condens. Matter **64**, 189 (1986).
2. M. K. Wu, J. Ashburn, C. J. Torng, P. H. Meng, L. Gao, Z. J. Huang, U. Q. Wang, and C. W. Chu, Phys. Rev. Lett. **58**, 908 (1987).
3. H. Maeda, Y. Tanaka, M. Fukutomi, and T. Asano, Jpn. J. Appl. Phys. **27**, L209 (1988).
4. A. Schilling, M. Cantoni, J. D. Guo, and H. R. Ott, Nature **363**, 56 (1993).
5. Y. Kamihara, T. Watanabe, M. Hirano, and H. Hosono, J. Am. Chem. Soc. **130**, 3296 (2008).
6. Z. A. Ren, W. Lu, J. Yang, W. Yi, X. L. Shen, Z. C. Li, G. C. Che, X. L. Dong, L. L. Sun, F. Zhou, and Z. X. Zhao, Chin. Phys. Lett. **25**, 2215 (2008).
7. M. Rotter, M. Tegel, and D. Johrendt, Phys. Rev. Lett. **101**, 107006 (2008).
8. X. C. Wang, Q. Q. Liu, Y. X. Lv, W. B. Gao, L. X. Yang, R. C. Yu, F. Y. Li, and C. Q. Jin, Solid State Commun. **148**, 538 (2008).
9. K. Ishida, Y. Nakai, and H. Hosono, J. Phys. Soc. Jpn. **78**, 062001 (2009).
10. Y. Mizuguchi and Y. Takano, J. Phys. Soc. Jpn. **79**, 102001 (2010).
11. Y. Tokura and T. Arima, Jpn. J. Appl. Phys. **29**, 2388 (1990).
12. Y. Mizuguchi, H. Fujihisa, Y. Gotoh, K. Suzuki, H. Usui, K. Kuroki, S. Demura, Y. Takano, H. Izawa, and O. Miura, Phys. Rev. B **86**, 220510 (2012).
13. Y. Mizuguchi, S. Demura, K. Deguchi, Y. Takano, H. Fujihisa, Y. Gotoh, H. Izawa, and O. Miura, J. Phys. Soc. Jpn. **81**, 114725 (2012).
14. J. Xing, S. Li, X. Ding, H. Yang, and H. H. Wen, Phys. Rev. B **86**, 214518 (2012).
15. S. Demura, K. Deguchi, Y. Mizuguchi, K. Sato, R. Honjyo, A. Yamashita, T. Yamaki, H. Hara, T. Watanabe, S. J. Denholme, M. Fujioka, H. Okazaki, T. Ozaki, O. Miura, T. Yamaguchi, H. Takeya, and Y. Takano, J. Phys. Soc. Jpn. **84**, 024709 (2015).
16. R. Jha, A. Kumar, S. K. Singh, and V. P. S. Awana, J. Supercond. Novel Magn. **26**, 499 (2013).
17. S. Demura, Y. Mizuguchi, K. Deguchi, H. Okazaki, H. Hara, T. Watanabe, S. J. Denholme, M. Fujioka, T. Ozaki, H. Fujihisa, Y. Gotoh, O. Miura, T. Yamaguchi, H. Takeya, and Y. Takano, J. Phys. Soc. Jpn. **82**, 033708 (2013).
18. H. Usui, K. Suzuki, and K. Kuroki, Phys. Rev. B **86**, 220501 (2012).
19. X. Wan, H. C. Ding, S. Y. Savrasov, and C. G. Duan, Phys. Rev. B **87**, 115124 (2013).
20. D. Yazici, K. Huang, B. D. White, A. H. Chang, A. J. Friedman, and M. B. Maple, Philos. Mag. **93**, 673 (2012).
21. S. Li, H. Yang, J. Tao, X. Ding, and H. H. Wen, Sci. China-Phys. Mech. Astron. **56**, 2019 (2013).
22. S. G. Tan, L. J. Li, Y. Liu, P. Tong, B. C. Zhao, W. J. Lu, and Y. P. Sun, Physica C **483**, 94 (2012).
23. S. K. Singh, A. Kumar, B. Gahtori, S. Kirtan, G. Sharma, S. Patnaik, and V. P. S. Awana, J. Am. Chem. Soc. **134**, 16504 (2012).
24. H. Kotegawa, Y. Tomita, H. Tou, H. Izawa, Y. Mizuguchi, O. Miura, S. Demura, K. Deguchi, and Y. Takano, J. Phys. Soc. Jpn. **81**, 103702 (2012).
25. H. Takatsu, Y. Mizuguchi, H. Izawa, O. Miura, and H. Kadowaki, J. Phys. Soc. Jpn. **81**, 125002 (2012).
26. C. I. Sathish, H. L. Feng, Y. Shi, and K. Yamaura, J. Phys. Soc. Jpn. **82**, 074703 (2013).
27. S. G. Tan, P. Tong, Y. Liu, W. J. Lu, L. J. Li, B. C. Zhao, and Y. P. Sun, Eur. Phys. J. B **85**, 414 (2012).
28. W. A. Phelan, D. C. Wallace, K. E. Arpino, J. R. Neilson, K. J. Livi, C. R. Seabourne, A. J. Scott, and T. M. McQueen, J. Am. Chem. Soc. **135**, 5372 (2013).
29. P. Srivatsava, Shruti, and S. Patnaik, Supercond. Sci. Technol. **27**, 055001 (2014).
30. Shruti, P. Srivastava, and S. Patnaik, J. Phys.: Condens. Matter **25**, 312202 (2013).
31. P. K. Biswas, A. Amato, C. Baines, R. Khasanov, H. Luetkens, H. Lei, C. Petrovic, and E. Morenzoni, Phys. Rev. B **88**,

224515 (2013).
32. Y. Yu, J. Shao, S. Tan, C. Zhang, and Y. Zhang, J. Phys. Soc. Jpn. **82**, 034718 (2013).
33. R. Jha and V. P. S. Awana, Physica C **498**, 45 (2014).
34. J. Shao, Z. Liu, X. Yao, L. Pi, S. Tan, C. Zhang, and Y. Zhang, Phys. Status Solidi RRL **8**, 845 (2014).
35. J. Shao, X. Yao, Z. Liu, L. Pi, S. Tan, C. Zhang, and Y. Zhang, Supercond. Sci. Technol. **28**, 015008 (2015).
36. X. Zhang, Y. Liu, G. Zhang, Y. Wang, H. Zhang, and F. Huang, ACS Appl. Mater. Interfaces **7**, 4442 (2015).
37. T. Okada, H. Ogino, J. Shomoyama, and K. Kishio, Appl. Phys. Express **8**, 023102 (2015).
38. D. Mancusi, F. Giubileo, Y. Mizuguchi, S. Pace, and M. Polichetti, Physica C **507**, 47 (2014).
39. T. Okada, H. Ogino, Jun-ichi Shimoyama, K. Kishio, N. Takeshita, N. Shirakawa, A. Iyo, and H. Eisaki, J. Phys. Soc. Jpn. **84**, 084703 (2015).
40. N. Takahashi, M. Nagao, A. Miura, S. Watauchi, K. Tadanaga, Y. Takano, and I. Tanaka, J. Ceram. Soc. Jpn. **126**, 591 (2018).
41. A. Miura, Y. Mizuguchi, T. Sugawara, Y. Wang, T. Takei, N. Kumada, E. Magome, C. Moriyoshi, Y. Kuroiwa, O. Miura, and K. Tadanaga, Inorg. Chem. **54**, 10462 (2015).
42. Z. Feng, X. Yin, Y. Cao, X. Peng, T. Gao, C. Yu, J. Chen, B. Kang, B. Lu, J. Guo, Q. Li, W. S. Tseng, Z. Ma, C. Jing, S. Cao, J. Zhang, and N. C. Yeh, Phys. Rev. B **94**, 064522 (2016).
43. X. Yin, Z. Feng, C. Yu, Q. Li, Y. Cao, B. Kang, B. Lu, J. Chen, T. Gao, X. Li, J. Guo, H. Chu, G. Wang, D. Deng, C. Jing, S. Cao, and J. Zhang, J. Supercond. Nov. Magn. **29**, 879 (2016).
44. G. C. Kim, M. Cheon, D. Ahmad, Y. S. Kwon, R. K. Ko, and Y. C. Kim, J. Phys. Soc. Jpn. **87**, 025003 (2018).
45. V. P. S. Awana, A. Kumar, R. Jha, S. Kumar, A. Pal, Shruti, J. Saha, and S. Patnaik, Solid State Commun. **157**, 31 (2013).
46. R. Jha, A. Kumar, S. K. Singh, and V. P. S. Awana, J. Appl. Phys. **113**, 056102 (2013).
47. H. Lei, K. Wang, M. Abeykoon, E. S. Bozin, and C. Petrovic, Inorg. Chem. **52**, 10685 (2013).
48. T. Zhou and Z. D. Wang, J. Supercond. Nov. Magn. **26**, 2735 (2013).
49. B. Li, Z. W. Xing, and G. Q. Huang, EPL **101**, 47002 (2013).
50. K. Deguchi, Y. Mizuguchi, S. Demura, H. Hara, T. Watanabe, S. J. Denholme, M. Fujioka, H. Okazaki, T. Ozaki, H. Takeya, T. Yamaguchi, O. Miura, and Y. Takano, EPL **101**, 17004 (2013).
51. X. Lin, X. Ni, B. Chen, X. Xu, X. Yang, J. Dai, Y. Li, X. Yang, Y. Luo, Q. Tao, G. Cao, and Z. Xu, Phys. Rev. B **87**, 020504 (2013).
52. D. Yazici, K. Huang, B. D. White, I. Jeon, V. W. Burnett, A. J. Friedman, I. K. Lum, M. Nallaiyan, S. Spagna, and M. B. Maple, Phys. Rev. B **87**, 174512 (2013).
53. J. Kajitani, K. Deguchi, A. Omachi, T. Hiroi, Y. Takano, H. Takatsu, H. Kadowaki, O. Miura, and Y. Mizuguchi, Solid State Commun. **181**, 1 (2014).
54. G. Kalai Selvan, M. Kanagaraj, S. Esakki Muthu, Rajveer Jha, V. P. S. Awana, and S. Arumugam, Phys. Status Solidi RRL **7**, 510 (2013).
55. C. T. Wolowiec, D. Yazici, B. D. White, K. Huang, and M. B. Maple, Phys. Rev. B **88**, 064503 (2013).
56. C. T. Wolowiec, B. D. White, I. Jeon, D. Yazici, K. Huang, and M. B. Maple, J. Phys.: Condens. Matter **25**, 422201 (2013).
57. M. Nagao, S. Demura, K. Deguchi, A. Miura, S. Watauchi, T. Takei, Y. Takano, N. Kumada, and I. Tanaka, J. Phys. Soc. Jpn. **82**, 113701 (2013).
58. J. Liu, D. Fang, Z. Wang, J. Xing, Z. Du, X. Zhu, H. Yang, and H. H. Wen, EPL **106**, 67002 (2014).
59. M. Nagao, A. Miura, S. Demura, K. Deguchi, S. Watauchi, T. Takei, Y. Takano, N. Kumada, and I. Tanaka, Solid State Commun. **178**, 33 (2014).
60. Y. Li, X. Lin, L. Li, N. Zhou, X. Xu, C. Cao, J. Dai, L. Zhang, Y. Luo, W. Jiao, Q. Tao, G. Cao, and Z. Xu, Supercond. Sci. Technol. **27**, 035009 (2014).
61. R. Jha and V. P. S. Awana, J. Supercond. Nov. Magn. **27**, 1 (2014).
62. A. Krzton-Maziopa, Z. Guguchia, E. Pomjakushina, V. Pomjakushin, R. Khasanov, H. Luetkens, P. Biswas, A. Amato, H. Keller, and K. Conder, J. Phys.: Condens. Matter **26**, 215702 (2014).

63. G. Lamura, T. Shiroka, P. Bonfà, S. Sanna, R. De Renzi, C. Baines, H. Luetkens, J. Kajitani, Y. Mizuguchi, O. Miura, K. Deguchi, S. Demura, Y. Takano, and M. Putti, Phys. Rev. B **88**, 180509 (2013).
64. J. Kajitani, A. Omachi, T. Hiroi, O. Miura, and Y. Mizuguchi, Physica C, **504**, 33 (2014).
65. H. Sakai, D. Kotajima, K. Saito, H. Wadati, Y. Wakisaka, M. Mizumaki, K. Nitta, Y. Tokura, and S. Ishiwata, J. Phys. Soc. Jpn. **83**, 014709 (2014).
66. R. Jha, H. Kishan, and V. P. S. Awana, J. Appl. Phys. **115**, 013902 (2014).
67. R. Jha and V. P. S. Awana, Mater. Res. Express **1**, 016002 (2014).
68. J. Kajitani, K. Deguchi, T. Hiroi, A. Omachi, S. Demura, Y. Takano, O. Miura, and Y. Mizuguchi, J. Phys. Soc. Jpn. **83**, 065002 (2014).
69. R. Jha, B. Tiwari, and V. P. S. Awana, J. Phys. Soc. Jpn. **83**, 063707 (2014).
70. L. K. Zeng, X. B. Wang, J. Ma, P. Richard, S. M. Nie, H. M. Weng, N. L. Wang, Z. Wang, T. Qian, and H. Ding, Phys. Rev. B **90**, 054512 (2014)
71. Z. R. Ye, H. F. Yang, D. W. Shen, J. Jiang, X. H. Niu, D. L. Feng, Y. P. Du, X. G. Wan, J. Z. Liu, X. Y. Zhu, H. H. Wen, and M. H. Jiang, Phys. Rev. B **90**, 045116 (2014)
72. J. Lee, M. B. Stone, A. Huq, T. Yildrim, G. Ehlers, Y. Mizuguchi, O. Miura, Y. Takano, K. Deguchi, S. Demura, and S. H. Lee, Phys. Rev. B **87**, 205134 (2013).
73. S. Nagira, J. Sonoyama, T. Wakita, M. Sunagawa, Y. Izumi, T. Muro, H. Kumigashira, M. Oshima, K. Deguchi, H. Okazaki, Y. Takano, O. Miura, Y. Mizuguchi, K. Suzuki, H. Usui, K. Kuroki, K. Okada, Y. Muraoka, and T. Yokoya, J. Phys. Soc. Jpn. **83**, 033703 (2014).
74. A. Miura, M. Nagao, T. Takei, S. Watauchi, I. Tanaka, and N. Kumada, J. Solid State Chem. **212**, 213 (2014).
75. T. Machida, Y. Fujisawa, M. Nagao, S. Demura, K. Deguchi, Y. Mizuguchi, Y. Takano, and H. Sakata, J. Phys. Soc. Jpn. **83**, 113701 (2014).
76. H. Chen, G. Zhang, T. Hu, G. Mu, W. Li, F. Huang, X. Xie, and M. Jiang, Inorg. Chem. **53**, 9 (2014).
77. T. Hiroi, J. Kajitani, A. Omachi, O. Miura, and Y. Mizuguchi, J. Supercond. Nov. Magn. **28**, 1149 (2015).
78. J. Kajitani, T. Hiroi, A. Omachi, O. Miura, and Y. Mizuguchi, J. Supercond. Nov. Magn. **28**, 1129 (2015).
79. X. C. Wang, D. Y. Chen, Q. Guo, J. Yu, B. B. Ruan, Q. G. Mu, G. F. Chen, and Z. A. Ren, arXiv:1404.7562.
80. I. Pallecchi, G. Lamura, M. Putti, J. Kajitani, Y. Mizuguchi, O. Miura, S. Demura, K. Deguchi, and Y. Takano, Phys. Rev. B **89**, 214513 (2014).
81. T. Tomita, M. Ebata, H. Soeda, H. Takahashi, H. Fujihisa, Y. Gotoh, Y. Mizuguchi, H. Izawa, O. Miura, S. Demura, K. Deguchi, and Y. Takano, J. Phys. Soc. Jpn. **83**, 063704 (2014).
82. Y. Mizuguchi, T. Hiroi, J. Kajitani, H. Takatsu, H. Kadowaki, and O. Miura, J. Phys. Soc. Jpn. **83**, 053704 (2014).
83. Y. Mizuguchi, A. Miyake, K. Akiba, M. Tokunaga, J. Kajitani, and O. Miura, Phys. Rev. B **89**, 174515 (2014).
84. A. Omachi, J. Kajitani, T. Hiroi, O. Miura, and Y. Mizuguchi, J. Appl. Phys. **115**, 083909 (2014).
85. A. Omachi, T. Hiroi, J. Kajitani, O. Miura, and Y. Mizuguchi, J. Phys.: Conf. Ser. **507**, 012033 (2014).
86. R. Jha, H. Kishan, and V. P. S. Awana, Solid State Commun. **194**, 6 (2014).
87. M. Tanaka, M. Nagao, Y. Matsushita, M. Fujioka, S. J. Denholme, T. Yamaguchi, H. Takeya, and Y. Takano, J. Solid State Chem. **219**, 168 (2014).
88. M. Nagao, M. Tanaka, S. Watauchi, I. Tanaka, and Y. Takano, J. Phys. Soc. Jpn. **83**, 114709 (2014).
89. T. Yildirim, Phys. Rev. B **87**, 020506 (2013).
90. I. R. Shein and A. L. Ivanovskii, JETP Lett. **96**, 769 (2012).
91. Y. Liang, X. Wu, W. F. Tsai, and J. Hu, Front. Phys. **9**, 194 (2014).
92. K. Suzuki, H. Usui, and K. Kuroki, Phys. Proc. **45**, 21 (2013).
93. G. Martins, A. Moreo, and E. Dagotto, Phys. Rev. B **87**, 081102 (2013).
94. H. Wang, Chin. Phys. Lett. **31**, 047802 (2014).
95. C. Morice, E. Artacho, S. E. Dutton, D. Molnar, H. J. Kim, and S. S. Saxena, J. Phys.: Condens. Matter

**27**, 135501 (2015).
96. H. Wang, Chin. Phys. Lett. **32**, 017801 (2015).
97. C. Morice, E. Artacho, S. E. Dutton, D. Molnar, H. J. Kim, and S. S. Saxena, J. Phys.: Condens. Matter **28**, 345504 (2016).
98. Y. Feng, H. C. Ding, Y. Du, X. Wan, B. Wang, S. Y. Savrasov, and C. G. Duan, J. Appl. Phys. **115**, 233901 (2014).
99. N. Benayad, M. Djermouni, and A. Zaoui, Computational Condensed Matter **1**, 19 (2014).
100. Y. Ma, Y. Dai, N. Yin, T. Jing, and B. Huang, J. Mater. Chem. C **2**, 8539 (2014).
101. T. Agatsuma and T. Hotta, J. Magn. Magn. Mater. **400**, 73 (2016).
102. M. Ochi, R. Akashi, and K. Kuroki, J. Phys. Soc. Jpn. **85**, 094705 (2016).
103. Q. Liu, X. Zhang, and A. Zunger, Phys. Rev. B **93**, 174119 (2016).
104. C. Morice, R. Akashi, T. Koretsune, S. S. Saxena, and R. Arita, Phys. Rev. B **95**, 180505 (2017).
105. K. Suzuki, H. Usui, K. Kuroki, and H. Ikeda, Phys. Rev. B **96**, 024513 (2017).
106. G. Wang, D. Wang, X. Shi, and Y. Peng, Mod. Phys. Lett. B **31**, 1750265 (2017).
107. X. Y. Dong, J. F. Wang, R. X. Zhang, W. H. Duan, B. F. Zhu, J. O. Sofo, and C. X. Liu, Nat. Commun. **6**, 8517 (2015).
108. Q. Liu, Y. Guo, and A. J. Freeman, Nano Lett. **13**, 5264 (2013).
109. S. Cobo-Lopez, M. S. Bahramy, R. Arita, A. Akbari, and I. Eremin, New J. Phys. **20**, 043029 (2018).
110. M. Ochi, H. Usui, and K. Kuroki, Phys. Rev. Applied **8**, 064020 (2017).
111. M. Fujioka, M. Nagao, S. J. Denholme, M. Tanaka, H. Takeya, T. Yamaguchi, and Y. Takano, Appl. Phys. Lett. **105**, 052601 (2014).
112. X. B. Wang, S. M. Nie, H. P. Wang, P. Zheng, P. Wang, T. Dong, H. M. Weng, and N. L. Wang, Phys. Rev. B **90**, 054507 (2014).
113. L. Jiao, Z. F. Weng, J. Z. Liu, J. L. Zhang, G. M. Pang, C. Y. Guo, F. Gao, X. Y. Zhu, H. H. Wen, and H. Q. Yuan, J. Phys.: Condens. Matter **27**, 225701 (2015).
114. R. Jha, B. Tiwari, and V. P. S. Awana, J. Appl. Phys. **117**, 013901 (2015).
115. L. Li, Y. Li, Y. Jin, H. Huang, B. Chen, X. Xu, J. Dai, L. Zhang, X. Yang, H. Zhai, G. Cao, and Z. Xu, Phys. Rev. B **91**, 014508 (2015).
116. J. Liu, S. Li, Y. Li, X. Zhu, and H. H. Wen, Phys. Rev. B **90**, 094507 (2014).
117. J. Shao, Z. Liu, X. Yao, L. Zhang, L. Pi, S. Tan, C. Zhang, and Y. Zhang, EPL **107**, 37006 (2014).
118. S. F. Wu, P. Richard, X. B. Wang, C. S. Lian, S. M. Nie, J. T. Wang, N. L. Wang, and H. Ding, Phys. Rev. B **90**, 054519 (2014).
119. R. Jha, B. Tiwari, and V. P. S. Awana, J. Phys. Soc. Jpn. **83**, 105001 (2014).
120. H. F. Zhai, Z. T. Tang, H. Jiang, K. Xu, K. Zhang, P. Zhang, J. K. Bao, Y. L. Sun, W. H. Jiao, I. Nowik, I. Felner, Y. K. Li, X. F. Xu, Q. Tao, C. M. Feng, Z. A. Xu, and G. H. Cao, Phys. Rev. B **90**, 064518 (2014).
121. M. Fujioka, M. Tanaka, S. J. Denholme, T. Yamaki, H. Takeya, T. Yamaguchi, and Y. Takano, EPL **108**, 47007 (2014).
122. I. Jeon, D. Yazici, B. D. White, A. J. Friedman, and M. B. Maple, Phys. Rev. B **90**, 054510 (2014).
123. Y. Mizuguchi, A. Omachi, Y. Goto, Y. Kamihara, M. Matoba, T. Hiroi, J. Kajitani, and O. Miura, J. Appl. Phys. **116**, 163915 (2014).
124. T. Sugimoto, B. Joseph, E. Paris, A. Iadecola, T. Mizokawa, S. Demura, Y. Mizuguchi, Y. Takano, and N. L. Saini, Phys. Rev. B **89**, 201117 (2014).
125. Y. Fang, D. Yazici, B. D. White, and M. B. Maple, Phys. Rev. B **91**, 064510 (2015).
126. G. S. Thakur, G. K. Selvan, Z. Haque, L. C. Gupta, S. L. Samal, S. Arumugam, and A. K. Ganguli, Inorg. Chem. **54**, 1076 (2015).
127. J. Kajitani, T. Hiroi, A. Omachi, O. Miura, and Y. Mizuguchi, J. Phys. Soc. Jpn. **84**, 044712 (2015).
128. T. Hiroi, J. Kajitani, A. Omachi, O. Miura, and Y. Mizuguchi, J. Phys. Soc. Jpn. **84**, 024723 (2015).
129. H. F. Zhai, P. Zhang, S. Q. Wu, C. Y. He, Z. T. Tang, H. Jiang, Y. L. Sun, J. K. Bao, I. Nowik, I. Felner, Y.

W. Zeng, Y. K. Li, X. F. Xu, Q. Tao, Z. A. Xu, and G. H. Cao, J. Am. Chem. Soc. **136**, 15386 (2014).
130. E. Paris, B. Joseph, A. Iadecola, T. Sugimoto, L. Olivi, S. Demura, Y. Mizuguchi, Y. Takano, T. Mizokawa, and N. L. Saini, J. Phys.: Condens. Matter **26**, 435701 (2014).
131. Y. Luo, H. F. Zhai, P. Zhang, Z. A. Xu, G. H. Cao, and J. D. Thompson, Phys. Rev. B **90**, 220510 (2014).
132. R. Higashinaka, T. Asano, T. Nakashima, K. Fushiya, Y. Mizuguchi, O. Miura, T. D. Matsuda, and Y. Aoki, J. Phys. Soc. Jpn. **84**, 023702 (2015).
133. M. Tanaka, T. Yamaki, Y. Matsushita, M. Fujioka, S. J. Denholme, T. Yamaguchi, H. Takeya, and Y. Takano, Appl. Phys. Lett. **106**, 112601 (2015).
134. K. Terashima, J. Sonoyama, T. Wakita, M. Sunagawa, K. Ono, H. Kumigashira, T. Muro, M. Nagao, S. Watauchi, I. Tanaka, H. Okazaki, Y. Takano, O. Miura, Y. Mizuguchi, H. Usui, K. Suzuki, K. Kuroki, Y. Muraoka, and T. Yokoya, Phys. Rev. B **90**, 220512 (2014).
135. J. Lee, S. Demura, M. B. Stone, K. Iida, G. Ehlers, C. R. dela Cruz, M. Matsuda, K. Deguchi, Y. Takano, Y. Mizuguchi, O. Miura, and S. H. Lee, Phys. Rev. B **90**, 224410 (2014).
136. A. Miura, M. Nagao, T. Takei, S. Watauchi, Y. Mizuguchi, Y. Takano, I. Tanaka, and N. Kumada, Cryst. Growth Des. **15**, 39 (2015).
137. Y. Tian, A. Zhang, J. Ji, J. Liu, X. Zhu, H. H. Wen, F. Jin, X. Ma, and Q. M. Zhang, Supercond. Sci. Technol. **29**, 015007 (2016).
138. R. Higashinaka, R. Miyazaki, Y. Mizuguchi, O. Miura, and Y. Aoki, J. Phys. Soc. Jpn. **83**, 075004 (2014).
139. R. Jha, and V. P. S. Awana, J. Supercond. Nov. Magn. **28**, L2229 (2015).
140. A. Athauda, J. Yang, S. Lee, Y. Mizuguchi, K. Deguchi, Y. Takano, O. Miura, and D. Louca, Phys. Rev. B **91**, 144112 (2014).
141. Y. Mizuguchi, A. Miura, J. Kajitani, T. Hiroi, O. Miura, K. Tadanaga, N. Kumada, E. Magome, C. Moriyoshi, and Y. Kuroiwa, Sci. Rep. **5**, 14968 (2015).
142. G. S. Thakur, R. Jha, Z. Haque, V. P. S. Awana, L. C. Gupta, and A. K. Ganguli, Supercond. Sci. Technol. **28**, 115010 (2015).
143. T. Sugimoto, D. Ootsuki, M. Takahashi, C. Morice, E. Artacho, S. S. Saxena, E. F. Schwier, M. Zheng, Y. Kojima, H. Iwasawa, K. Shimada, M. Arita, H. Namatame, M. Taniguchi, N. L. Saini, T. Asano, T. Nakajima, R. Higashinaka, T. D. Matsuda, Y. Aoki, and T. Mizokawa, Phys. Rev. B **92**, 041113 (2015).
144. C. Y. Guo, Y. Chen, M. Smidman, S. A. Chen, W. B. Jiang, H. F. Zhai, Y. F. Wang, G. H. Cao, J. M. Chen, X. Lu, and H. Q. Yuan, Phys. Rev. B **91**, 214512 (2015).
145. H. F. Zhai, P. Zhang, Z. T. Tang, J. K. Bao, H. Jiang, C. M. Feng, Z. A. Xu, and G. H. Cao, J. Phys.: Condens. Matter **27**, 385701 (2015).
146. K. Suzuki, M. Tanaka, S. J. Denholme, M. Fujioka, T. Yamaguchi, H. Takeya, and Y. Takano, J. Phys. Soc. Jpn. **84**, 115003 (2015).
147. M. Nagao, A. Miura, S. Watauchi, Y. Takano, and I. Tanaka, Jpn. J. Appl. Phys. **54**, 083101 (2015).
148. Y. Goto, J. Kajitani, Y. Mizuguchi, Y. Kamihara, and M. Matoba, J. Phys. Soc. Jpn. **84**, 085003 (2015).
149. Y. Fang, D. Yazici, B. D. White, and M. B. Maple, Phys. Rev. B **92**, 094507 (2015).
150. G. K. Selvan, G. Thakur, K. Manikandan, A. Banerjee, Z. Haque, L. C. Gupta, A. Ganguli, and S. Arumugam, J. Phys. D: Appl. Phys. **49**, 275002 (2016).
151. E. Uesugi, S. Nishiyama, H. Akiyoshi, H. Goto, Y. Koike, K. Yamada, and Y. Kubozono, Adv. Electron. Mater. **1**, 1500085 (2015).
152. E. Uesugi, S. Nishiyama, H. Goto, H. Ota, and Y. Kubozono, Appl. Phys. Lett. **109**, 252601 (2016).
153. G. Kalai Selvan, G. S. Thakur, K. Manikandan, Y. Uwatoko, A. K. Ganguli, and S. Arumugam, J. Phys. Soc. Jpn. **84**, 124701 (2015).
154. A. Miura, Y. Mizuguchi, T. Takei, N. Kumada, E. Magome, C. Moriyoshi, Y. Kuroiwa, and K. Tadanaga, Solid State Commun. **227**, 19 (2016).
155. Y. Mizuguchi, E. Paris, T. Sugimoto, A. Iadecola, J. Kajitani, O. Miura, T. Mizokawa, and N. L. Saini, Phys. Chem. Chem. Phys. **17**, 22090 (2015).

156. A. Nishida, O. Miura, C. H. Lee, and Y. Mizuguchi, Appl. Phys. Express **8**, 111801 (2015).
157. P. Zhang, H. F. Zhai, Z. J. Tang, L. Li, Y. K. Li, Q. Chen, J. Chen, Z. Wang, C. M. Feng, G. H. Cao, and Z. A. Xu, EPL **111**, 27002 (2015).
158. S. Demura, Y. Fujisawa, S. Otsuki, R. Ishio, Y. Takano, and H. Sakata, Solid State Commun. **223**, 40 (2015).
159. Y. Mizuguchi, T. Hiroi, and O. Miura, J. Phys.: Conf. Ser. **683**, 012001 (2016).
160. L. Li, D. Parker, P. Babkevich, L. Yang, H. M. Ronnow, and A. S. Sefat, Phys. Rev. B **91**, 104511 (2015).
161. R. Sagayama, H. Sagayama, R. Kumai, Y. Murakami, T. Asano, J. Kajitani, R. Higashinaka, T. D. Matsuda, and Y. Aoki, J. Phys. Soc. Jpn. **84**, 123703 (2015).
162. A. Nishida, H. Nishiate, C. H. Lee, O. Miura, and Y. Mizuguchi, J. Phys. Soc. Jpn. **85**, 074702 (2016).
163. T. Yamashita, Y. Tokiwa, D. Terazawa, M. Nagao, S. Watauchi, I. Tanaka, T. Terashima, and Y. Matsuda, J. Phys. Soc. Jpn. **85**, 073707 (2016).
164. A. Athauda, C. Hoffman, Y. Ren, S. Aswartham, J. Terzic, G. Cao, X. Zhu, and D. Louca, J. Phys. Soc. Jpn. **86**, 054701 (2017).
165. M. Aslam, A. Paul, G. S. Thakur, S. Gayen, R. Kumar, A. Singh, S. Das, A. K. Ganguli, U. V. Waghmare, and G. Sheet, J. Phys.: Condens. Matter **28**, 195701 (2016).
166. Y. Mizuguchi, A. Miura, A. Nishida, O. Miura, K. Tadanaga, N. Kumada, C. H. Lee, E. Magome, C. Moriyoshi, and Y. Kuroiwa, J. Appl. Phys. **119**, 155103 (2016).
167. T. Sugimoto, D. Ootsuki, E. Paris, A. Iadecola, M. Salome, E. F. Schwier, H. Iwasawa, K. Shimada, T. Asano, R. Higashinaka, T. D. Matsuda, Y. Aoki, N. L. Saini, and T. Mizokawa, Phys. Rev. B **94**, 081106 (2016).
168. M. Aslam, S. Gayen, A. Singh, M. Tanaka, T. Yamaki, Y. Takano, and G. Sheet, Solid State Commun. **264**, 26 (2017).
169. L. Li, Y. Xiang, Y. Chen, W. Jiao, C. Zhang, L. Zhang, J. Dai, and Y. Li, Supercond. Sci. Technol. **29**, 04LT03, (2016).
170. X. Zhou, Q. Liu, J. A. Waugh, H. Li, T. Nummy, X. Zhang, X. Zhu, G. Cao, A. Zunger, and D. S. Dessau, Phys. Rev. B **95**, 075118 (2017).
171. M. Nagao, A. Miura, I. Ueta, S. Watauchi, and I. Tanaka, Solid State Commun. **245**, 11 (2016).
172. Y. Y. Chin, H. J. Lin, Z. Hu, M. Nagao, Y. Du, X. Wan, B. J. Su, L. Y. Jang, T. S. Chan, H. Y. Chen, Y. L. Soo, and C. T. Chen, Phys. Rev. B **94**, 035150 (2016).
173. J. D. Querales-Flores, C. I. Ventura, R. Citro, and J. J. Rodriguez-Nunez, Physica B **488**, 32 (2016).
174. G. Jinno, R. Jha, A. Yamada, R. Higashinaka, T. D. Matsuda, Y. Aoki, M. Nagao, O. Miura, and Y. Mizuguchi, J. Phys. Soc. Jpn. **85**, 124708 (2016).
175. S. L. Wu, Z. A. Sun, F. K. Chiang, C. Ma, H. F. Tian, R. X. Zhang, B. Zhang, J. Q. Li, and H. X. Yang, Solid State Commun. **205**, 14 (2015).
176. Y. Fang, C. T. Wolowiec, A. J. Breindel, D. Yazici, P. C. Ho, and M. B. Maple, Supercond. Sci. Technol. **30**, 115004 (2017).
177. J. Zhang, K. Huang, Z. F. Ding, D. E. MacLaughlin, O. O. Bernal, P. C. Ho, C. Tan, X. Liu, D. Yazici, M. B. Maple, and L. Shu, Phys. Rev. B **94**, 224502 (2016).
178. M. Kannan, G. K. Selvan, Z. Haque, G. S. Thakur, B. Wang, K. Ishigaki, Y. Uwatoko, L. C. Gupta, A. K. Ganguli, and S. Arumugam, Supercond. Sci. Technol. **30**, 115011 (2017).
179. Z. Haque, G. S. Thakur, R. Parthasarathy, B. Gerke, T. Block, L. Heletta, R. Pottgen, A. G. Joshi, G. K. Selvan, S. Arumugam, L. C. Gupta, and A. K. Ganguli, Inorg. Chem. **56**, 3182 (2017).
180. Z. Haque, G. S. Thakur, R. Pottgen, G. K. Selvan, R. Parthasarathy, S. Arumugam, L. C. Gupta, and A. K. Ganguli, Inorg. Chem. **57**, 37 (2018).
181. J. Cheng, H. F. Zhai, Y. Wang, W. Xu, S. Liu, and G. H. Cao, Sci. Rep. **6**, 37394 (2016).
182. J. Zhan, L. Li, T. Wang, J. Wang, Y. Chen, L. Zhang, J. Shen, P. Li, and Y. Li, J. Supercond. Nov. Magn. **30**, 305 (2016).

183. K. Nagasaka, G. Jinno, O. Miura, A. Miura, C. Moriyoshi, Y. Kuroiwa, and Y. Mizuguchi, J. Phys.: Conf. Ser. **871**, 012007 (2017).
184. K. Nagasaka, A. Nishida, R. Jha, J. Kajitani, O. Miura, R. Higashinaka, T. D. Matsuda, Y. Aoki, A. Miura, C. Moriyoshi, Y. Kuroiwa, H. Usui, K. Kuroki, and Y. Mizuguchi, J. Phys. Soc. Jpn. **86**, 074701 (2017).
185. Y. Mizuguchi, E. Paris, T. Wakita, G. Jinno, A. Puri, K. Terashima, B. Joseph, O. Miura, T. Yokoya, and N. L. Saini, Phys. Rev. B **95**, 064515 (2017).
186. E. Paris, Y. Mizuguchi, M. Y. Hacisalihoglu, T. Hiroi, B. Joseph, G. Aquilanti, O. Miura, T. Mizokawa, and N. L. Saini, J. Phys.: Condens. Matter **29**, 145603 (2017).
187. Y. Ota, K. Okazaki, H. Q. Yamamoto, T. Yamamoto, S. Watanabe, C. Chen, M. Nagao, S. Watauchi, I. Tanaka, Y. Takano, and S. Shin, Phys. Rev. Lett. **118**, 167002 (2017).
188. M. Nagao, M. Tanaka, S. Watauchi, Y. Takano, and I. Tanaka, Solid State Commun. **261**, 32 (2017).
189. M. Tanaka, M. Nagao, R. Matsumoto, N. Kataoka, I. Ueta, H. Tanaka, S. Watauchi, I. Tanaka, and Y. Takano, J. Alloys Compd. **722**, 467 (2017).
190. A. Bhattacharyya, D. T. Adroja, A. D. Hillier, R. Jha, V. P. S. Awana, and A. M. Strydom, J. Phys.: Condens. Matter **29**, 265602 (2017).
191. Y. Goto, R. Sogabe, and Y. Mizuguchi, J. Phys. Soc. Jpn. **86**, 104712 (2017).
192. A. Athauda, Y. Mizuguchi, M. Nagao, J. Neuefeind, and D. Louca, J. Phys. Soc. Jpn. **86**, 124718 (2017).
193. Y. L. Sun, A. Ablimit, H. F. Zhai, J. K. Bao, Z. T. Tang, X. B. Wang, N. L. Wang, C. M. Feng, and G. H. Cao, Inorg. Chem. **53**, 11125 (2014).
194. Y. Mizuguchi, Y. Hijikata, T. Abe, C. Moriyoshi, Y. Kuroiwa, Y. Goto, A. Miura, S. Lee, S. Torii, T. Kamiyama, C. H. Lee, M. Ochi, and K. Kuroki, EPL **119**, 26002 (2017).
195. K. Hoshi, Y. Goto, and Y. Mizuguchi, Phys. Rev. B **97**, 094509 (2018).
196. R. Sogabe, Y. Goto, A. Nishida, T. Katase, and Y. Mizuguchi, EPL **122**, 17004 (2018).
197. E. Paris, T. Sugimoto, T. Wakita, A. Barinov, K. Terashima, V. Kandyba, O. Proux, J. Kajitani, R. Higashinaka, T. D. Matsuda, Y. Aoki, T. Yokoya, T. Mizokawa, and N. L. Saini, Phys. Rev. B **95**, 035152 (2017).
198. J. Cheng, Y. Yang, Y. Li, P. Dong, Y. Wang, and S. Liu, J. Phys. Chem. C **121**, 8525 (2017).
199. A. M. Nikitin, V. Grinenko, R. Sarkar, J. C. Orain, M. V. Salis, J. Henke, Y. K. Huang, H. H. Klauss, A. Amato, and A. de Visser, Sci. Rep. **7**, 17370 (2017).
200. S. Demura, N. Ishida, Y. Fujisawa, and H. Sakata, J. Phys. Soc. Jpn. **86**, 113701 (2017).
201. N. Kase, Y. Terui, T. Nakano, and N. Takeda, Phys. Rev. B **96**, 214506 (2017).
202. Y. Terui, N. Kase, T. Nakano, and N. Takeda, J. Phys.: Conf. Ser. **871**, 012005 (2017).
203. Y. C. Chan, K. Y. Yip, Y. W. Cheung, Y. T. Chan, Q. Niu, J. Kajitani, R. Higashinaka, T. D. Matsuda, Y. Yanase, Y. Aoki, K. T. Lai, and S. K. Goh, Phys. Rev. B **97**, 104509 (2018).
204. S. Otsuki, S. Demura, Y. Sakai, Y. Fujisawa, and H. Sakata, Solid State Commun. **270**, 17 (2017).
205. Y. Hijikata, T. Abe, C. Moriyoshi, Y. Kuroiwa, Y. Goto, A. Miura, K. Tadanaga, Y. Wang, O. Miura, and Y. Mizuguchi, J. Phys. Soc. Jpn. **86**, 124802 (2017).
206. Y. Mizuguchi, K. Hoshi, Y. Goto, A. Miura, K. Tadanaga, C. Moriyoshi, and Y. Kuroiwa, J. Phys. Soc. Jpn. **87**, 023704 (2018).
207. C. H. Lee, A. Nishida, T. Hasegawa, H. Nishiate, H. Kunioka, S. Ohira-Kawamura, M. Nakamura, K. Nakajima, and Y. Mizuguchi, Appl. Phys. Lett. **112**, 023903 (2018).
208. Y. Fang, D. Yazici, I. Jeon, and M. B. Maple, Phys. Rev. B **96**, 214505 (2017).
209. T. Sugimoto, E. Paris, T. Wakita, K. Terashima, T. Yokoya, A. Barinov, J. Kajitani, R. Higashinaka, T. D. Matsuda, Y. Aoki, T. Mizokawa, and N. L. Saini, Sci. Rep. **8**, 2011 (2018).
210. F. Giubileo, F. Romeo, A. Di Bartolomeo, Y. Mizuguchi, and P. Romano, J. Phys. Chem. Solids **118**, 192 (2018).
211. J. Cheng, P. Zhang, P. Dong, X. Wang, Y. Wang, X. Li, S. Liu, Y. Li, and Z. Xu, J. Alloys Compd. **743**, 547 (2018).

212. R. Sogabe, Y. Goto, and Y. Mizuguchi, Appl. Phys. Express **11**, 053102 (2018).
213. Y. Goto, A. Miura, R. Sakagami, Y. Kamihara, C. Moriyoshi, Y. Kuroiwa, and Y. Mizuguchi, J. Phys. Soc. Jpn. **87**, 074703 (2018).
214. R. Jha, Y. Goto, R. Higashinaka, T. D. Matsuda, Y. Aoki, and Y. Mizuguchi, J. Phys. Soc. Jpn. **87**, 083704 (2018).
215. M. Nagao, M. Tanaka, R. Matsumoto, H. Tanaka, S. Watauchi, Y. Takano, and I. Tanaka, Cryst. Growth Des. **16**, 3037 (2016).
216. A. Miura, M. Nagao, Y. Goto, Y. Mizuguchi, T. D. Matsuda, Y. Aoki, C. Moriyoshi, Y. Kuroiwa, Y. Takano, S. Watauchi, I. Tanaka, N. C. Rosero-Navarro, and K. Tadanaga, Inorg. Chem. **57**, 5364 (2018).
217. S. Ishii, Y. Hongu, A. Miura, and M. Nagao, Appl. Phys. Express **9**, 063101 (2016).
218. A. Miura, T. Oshima, K. Maeda, Y. Mizuguchi, C. Moriyoshi, Y. Kuroiwa, Y. Meng, X. D. Wen, M. Nagao, M. Higuchi, and K. Tadanaga, J. Mater. Chem. A **5**, 14270 (2017).
219. R. Sogabe, Y. Goto, T. Abe, C. Moriyoshi, Y. Kuroiwa, A. Miura, K. Tadanaga, and Y. Mizuguchi, arXiv:1808.04090.
220. K. Momma and F. Izumi, J. Appl. Crystallogr. **41**, 653 (2008).
221. M. Nagao, Nov. Supercond. Mater. **1**, 64 (2015).
222. E. Paris, Y. Mizuguchi, T. Wakita, K. Terashima, T. Yokoya, T. Mizokawa, and N. L. Saini, J. Phys.: Condens. Matter. **30**, 455703 (2018).